\documentclass[10pt,journal,compsoc]{IEEEtran}
\ifCLASSOPTIONcompsoc
  \usepackage[nocompress]{cite}
\else
  \usepackage{cite}
\fi
\ifCLASSINFOpdf
\else
\fi
\usepackage{balance}  
\usepackage{graphics} 
\usepackage{txfonts}
\usepackage{times}    
\usepackage[pdftex]{hyperref}
\usepackage{color}
\usepackage{textcomp}
\usepackage{booktabs}
\usepackage{ccicons}
\usepackage{upgreek}

\usepackage{cite}
\usepackage{url}
\usepackage{fancybox}
\usepackage{multirow}
\usepackage{flushend}
\usepackage{booktabs}
\usepackage{tabularx}
\usepackage{comment}
\usepackage{array}
\usepackage[flushleft]{threeparttable}
\usepackage{mdframed}
\graphicspath{{}{images/}{dia/}}
\DeclareGraphicsExtensions{.pdf,.png}

\usepackage{listings}
\usepackage{courier}
\usepackage{hyperref}

\usepackage{xpatch}

\xpatchcmd{\refstepcounter}{%
  \stepcounter{#1}%
}{%
  \stepcounter{#1}%
  \xdef\scr{\number\value{#1}}%
}{\typeout{success}}{\typeout{failure}}

\usepackage{tikz}
\definecolor{1c1}{RGB}{188,162,6}
\definecolor{1c2}{RGB}{137,129,80}
\definecolor{1c3}{RGB}{239,167,31}
\definecolor{1c4}{RGB}{88,194,241}
\definecolor{1c5}{RGB}{6,180,188}

\tikzset{mynode/.style={draw=white,solid,circle,fill=green,inner sep=1pt, thick,
text=black}}
\tikzset{arrow line/.style={dashed, line width= 2.5pt, color=#1}}

\def\bf{\textbf}

\def\fig {Figure~}

\def\tbl {Table~}

\def\sec {Section~}
\def\secs {Sections~}

\def\it{\textit}

\def\tt{\mct}

\newcommand{\mct}[1]{{\footnotesize {\texttt {#1}}}}

\usepackage{paralist}

\newcommand{\nd}{\vspace{1mm}\noindent}
\usepackage[small,bf]{caption}
\usepackage{tikz}
\newcommand*\circled[1]{\tikz[baseline=(char.base)]{
            \node[shape=circle,draw,inner sep=1pt] (char) {#1};}}

\lstdefinestyle{inlinecode}{basicstyle={\ttfamily\scriptsize\bfseries}}

\newcommand{\urls}[1]{{\scriptsize\url{#1}}}
\usepackage{tcolorbox}
\newcommand{\emt}[1]{\emph{``#1''}}

\usepackage{paralist}
\usepackage[outercaption]{sidecap}
\usepackage [autostyle, english = american]{csquotes}
\MakeOuterQuote{"}
\newcounter{scn}
\usepackage[shortlabels]{enumitem}
\usepackage{bchart}
\begin{document}
\title{Reputation Manipulation in Stack Overflow}
\author{Gias Uddin\thanks{Gias Uddin (Corresponding Author), SWAT Lab, Polytechnique Montr\'{e}al. Email: gias.uddin@polymtl.ca}, 
Iren Mazloomzadeh\thanks{Iren Mazloomzadeh, Shiraz University. Email: i.mazloomzadeh17@gmail.com}, 
Foutse Khomh\thanks{Foutse Khomh, SWAT Lab, Polytechnique Montr\'{e}al. Email: foutse.khomh@polymtl.ca}, 
and Ashkan Sami\thanks{Ashkan Sami, Shiraz University. Email: sami@shirazu.ac.ir}}

\markboth{IEEE TRANSACTIONS ON SOFTWARE ENGINEERING,~Vol.~X, No.~X,
Month~2020}%
{Uddin \MakeLowercase{\textit{et al.}}: Reputation Manipulation in Stack Overflow}

\IEEEtitleabstractindextext{%
\begin{abstract}
The success of a crowd-sourced developer forum such as Stack Overflow depends on the quality of the contributions of its users. 
To ensure quality in shared knowledge and user participation, the incentive system in Stack Overflow awards users with reputation scores. 
While such mechanism can help improve user participation, the 
decentralized nature of the forum may make the incentive system prone to manipulation by users driven by personal agenda. This paper offers, for the first time, a 
comprehensive overview of the types of reputation manipulation scenarios that are exercised in Stack Overflow and the prevalence of such reputation gamers (i.e., frauds) in the platform.  
To understand the different types of reputation fraud scenarios, we conducted a qualitative study of 1697 posts from meta Stack Exchange sites 
where those scenarios are discussed. We found six different types of fraud scenarios, such as voting rings where frauds form communities and upvote each other repeatedly on similar posts. 
We then sought to automatically identify those suspicious reputations in Stack Overflow. 
By analyzing how such suspicious users are reported in meta Stack Exchange sites by other users, we developed two algorithms. For example, a suspicious user 
may more likely to work in a small isolated community and may show an unusual jump in his reputation score in a short time.
The first algorithm identifies isolated/semi-isolated communities where suspicious reputation frauds mostly collaborate with each other. 
The second algorithm looks for sudden unusual big jumps in the reputation scores of users. 
Each algorithm outputs a list of users as potential reputation frauds. We evaluated the performance of our algorithms by examining the 
reputation history dashboard of Stack Overflow users from the Stack Overflow website. We observed that around 60-80\% of users that are considered to be frauds by our algorithms got their reputation scores removed by Stack Overflow due to suspicious activities. 
Despite that many of those frauds are still active in Stack Overflow, and many of them got their scores removed multiple times. 
Stack Overflow may consider improving their fraud detection by considering our developed algorithms. Research in software engineering that 
utilize Stack Overflow data can benefit from the exclusion from posts created by fraud users to ensure good and authentic contents are used, e.g., to create API documentation 
or to suggest answers to an unanswered question, and so on.  

\end{abstract}
}

%


\maketitle

\IEEEdisplaynontitleabstractindextext

%
\IEEEpeerreviewmaketitle

\section{Introduction}\label{sec:introduction}
The online crowd-sourced developer forums, such as Stack Overflow, are becoming increasingly 
important to software developers and practitioners due to the shortcomings in traditional 
and official knowledge resources (e.g., documentation)~\cite{Ponzanelli-PrompterRecommender-EMSE2014,Robillard-FieldStudyAPILearningObstacles-SpringerEmpirical2011a,Uddin-HowAPIDocumentationFails-IEEESW2015}. 
The adoption, growth, and continued success of 
Stack Overflow depend on two major factors, the participation of users and the
quality of the shared
knowledge~\cite{Bagozzi-ParticipationLinuxUserGroups-JMC2006,Lakhani-FreeUserToUserAssistance-JRP2003,Parnin-CrowdDoc-TechReport2012}.
Unlike traditional knowledge sharing venues that were paid events, the developers participating in crowd-sourced forums are driven by the needs to learn from each other (which is often free), to become part of the
community, as well as the need to be recognized by peers or hiring organizations~\cite{website:Bethany-sohiring-blog2013,Bird-CollaborationOSS-ICSM2011a,Bird-LatentSocialOSS-FSE2008a}. 

The recognition of developers in Stack Overflow 
is aided through the \it{reputation/incentive} system~\cite{website:SO-ReputationSystem-2019,Vasilescu-SocialQAKnowledgeSharing-CSCW2014}. 
In Stack Overflow, users with higher reputations are considered as experts. Users with high reputations in Stack Overflow can significantly influence the overall knowledge-sharing process and content quality. \revision{Privileges determine what you're allowed to do on Stack Overflow; users with high reputation scores can gain more privileges\footnote{https://stackoverflow.com/help/privileges}.} In essence, the reputation rating of each user is a composite of the scores received by sharing knowledge (i.e., asking a question/answering to a question), the awarded badges and the role offered in Stack Overflow. The atomic unit of analysis of a user's 
reputation is the score, i.e., 
the votes he receives~\cite{website:SO-ReputationSystem-2019}. A user can get upvotes for sharing useful contents and downvotes or no votes for sharing useless/harmful contents.
     
The concept of user participation through the incentive system arises from \it{gamification}, which consists in using elements from game design to a 
non-game context (i.e., knowledge sharing)~\cite{Deterding-GamificationDesigningForMotivation-Interactions2012,Deterding-GamificationUsingDesignElemetsInNonGaming-CHI2011}. 
Gamification is widely used in online platforms to motivate the users to contribute more~\cite{Anderson-DiscoveringValueCommunityActivitySO-SIGKDD2012,Antin-BadgeInSocialMedia-CHI2011,Cavusoglu-GamificationMotivateVoluntaryCont-CSCW2015}. 
Psychological theories support the success of gamification to increase participation~\cite{Vassileva-MotivatingParticipationSocialComputing-UMI2012}. 
Indeed, interviews of contributors to Apache open source projects have shown that 
people who wish to be known as `experts' are more active in the community~\cite{Lakhani-FreeUserToUserAssistance-JRP2003}.
\begin{figure*}[t]
\centering
	\centering
   	\includegraphics[width=\textwidth,height=\textheight,keepaspectratio]{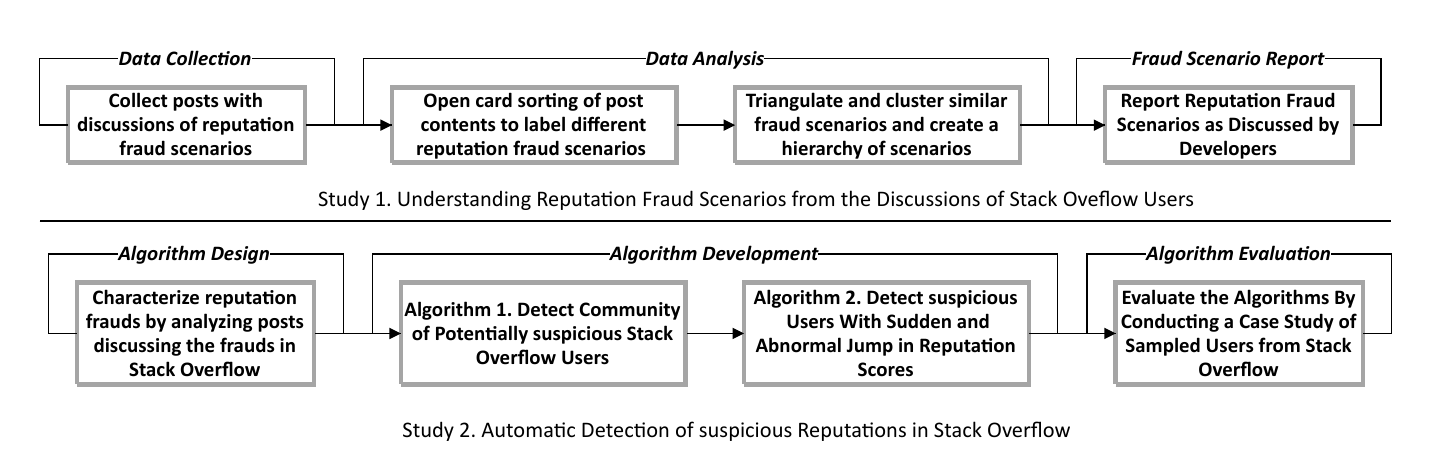}
   	\caption{The major steps to understand reputation fraud scenarios (Study 1) and to detect suspicious reputations (Study 2) in Stack Overflow}
   	 \label{fig:MethodologyOverall}
\end{figure*}

The danger of employing an incentive system in a crowd-sourced platform comes from the privileges that are associated with reputation and the incentive mechanisms. In the case of Stack Overflow,  
users have the power to vote for the shared contents and 
high reputation scores correlate with increased trust among users, which could incite unscrupulous users to fraud. 
For example, in a study of the reputation system of eBay, Resnick and 
Zeckhauser~\cite{Resnick-TrustEbayRepSystem-BookChapter2002} observed that %
when eBay decided to remove the reputation history of all sellers, the sellers with previously low reputation scores portrayed themselves as high quality sellers and manipulated the buyers~\cite{Ye-StrategicBehaviorReputationSystemEbay-MIS2014}.  
Similar deception is possible on Stack Overflow. Unscrupulous users can team up to manipulate the reputation system, by promoting each others posts to gain high scores in a short time (i.e., forming a \it{ring}). 
\acceptance{ In a 2016 Stack Overflow post, a highly reputed user (among the top 2\%), cautioned about the prevalence of such voting rings ~\cite{website:meta.stackoverflow-comment339072_322114}.
\begingroup
\addtolength\leftmargini{-0.1in}
\begin{quote}
	\it{"There has been a spike in such voting rings recently :("}
\end{quote}
\endgroup
\nd One developer was unaware of such abuse~\cite{website:meta.stackoverflow-comment339072_322114}. }
\begingroup
\addtolength\leftmargini{-0.1in}
\begin{quote}
	\it{"Personally I'm surprised we managed to go nearly 6 whole years without having to deal with anything like this."}
\end{quote}
\endgroup
\nd Another developer was worried that this could be an ongoing issue.
\begingroup
\addtolength\leftmargini{-0.1in}
\begin{quote}
	\it{"@BoltClock Here's a scary thought: What if we haven't, and its just been 6 years since we were able to catch one?"}
\end{quote}
\endgroup
\nd \acceptance{In another 2016 Stack Overflow post ~\cite{website:meta.stackexchange-answer-272729}, a moderator not only points to a voting ring fraud but also takes decisive actions by deleting the users' account and the corresponding votes related to the user. This is an indication that fraud and how it is formed has been a concern for Stack Overflow for the last few years.}
\begingroup
\addtolength\leftmargini{-0.1in}
\begin{quote}
\emt{I deleted your account on November 21. You and several others all clearly conspired to defraud the voting system by participating in a voting ring that was propping up low-quality or incorrect content. This voting coordination came to the attention of moderators through community flags or through our standard tools.
As your account had contributed little of value beyond the coordinated voting, I felt it safe to delete it in order to invalidate these votes. The friends of yours with more significant contributions were strongly warned about this and their votes manually invalidated by Stack Exchange employees.
In the future, I highly recommend not coordinating votes between yourself and your friends in an effort to gain an unfair advantage over others.}
\end{quote}
\endgroup
Indeed, manipulation of the reputation mechanism can introduce low quality 
contents in Stack Overflow. According to one user in Stack Overflow, \emt{I remember at the beginning of StackOverflow learning a lot by looking up
at the questions. Now... well, finding a good question in the most popular tags is just hard. I keep trudging through crap}\footnote{https://meta.stackoverflow.com/a/256304/}. In 
our previous study, a common concern raised by Stack Overflow developers was the inherent bias in the shared contents 
and the lack of proper mechanism to measure the trustworthiness of the contents and the users~\cite{Uddin-SurveyOpinion-TSE2019}.

\revision{Although evidence shows the prevalence of manipulations of the reputation system in Stack Overflow more than 10 years ago, we are not aware of any study, based on the best of our knowledge, that has investigated it in any form} ~\cite{website:meta.stackexchange-answer-53807}. 
Clearly, such a study is warranted given recent focus on the quality of 
contents shared in Stack Overflow~\cite{Zhang-AreCodeExamplesInForumReliable-ICSE2018,Ya-WantAGoodAnswer-Arxiv2013,Hudson-TriggerForClarificationRequestsForums-VLHCC2015,Masud-InsightUnresolvedQuestionsSO-MSR2015,Agichtein-FindingHighQualityContents-WSDM2008,Mondal-SOIssueReproducability-MSR2019}, 
the suite of tools and techniques developed to recommend quality 
posts~\cite{Ponzanelli-ClassifyQualityForumQuestion-QSIC2014,Ponzanelli-ImproveLowQualityPostDetect-ICSME2014,Ya-DetectHighQualityPosts-JIS2015,
Harper-PredictorAnswerQuality-CHI2008,Li-AnswerQualityPredictions-WWW2015,Calefato-HowToAskForTechnicalHelp-IST2018}, 
to produce software documentation~\cite{Subramanian-LiveAPIDocumentation-ICSE2014}, and to collect and summarize developers' reviews~\cite{Uddin-OpinionValue-TSE2019,Uddin-OpinerReviewAlgo-ASE2017,Uddin-OpinerReviewAlgo-ASE2017}. 
An understanding of reputation manipulation mechanisms in Stack Overflow can, for example, guide research in software engineering to filter contents shared by suspicious reputations, tell hiring agencies whom to or not to consider, and inform other users in Stack Overflow about the validity and trustworthiness of shared contents and knowledge providers.

\revision{The implications of this work for software engineering are manifold. Firstly, numerous research endeavors aiming to harness information from Stack Overflow to aid software development rely on reputation to pinpoint experts on the platform. Consequently, gaming practices are likely to introduce noise into this data, potentially compromising the validity of these findings. Moreover, reputation on Stack Overflow could impact promotions or job prospects, with gaming potentially resulting in inappropriate offers or advancements. Finally, we have reached out to the moderators and Stack Overflow team to inquire whether our research could enhance their fraud detection system. They expressed appreciation for our efforts and stressed that any form of gaming on Stack Overflow would result in low-quality posts, and the migration of such posts into real-world projects could yield low-quality applications.}

With a view to understand the possible types and prevalence of reputation manipulation in Stack Overflow, we report 
the results from two empirical studies. The studies offer insights into how exploitation may happen based on diverse reputation fraud scenarios 
and whether such suspicious reputations can be identified. \fig\ref{fig:MethodologyOverall} outlines the research methodology we followed to conduct the two studies. 

\nd\bf{$\bullet$ Study 1. Understanding Reputation Fraud Scenarios.} Our focus was to learn about the diverse reputation 
manipulation scenarios that fraud users in Stack Overflow are currently exercising. In the absence of any previous research or available data 
to guide us through this understanding, we decided to learn from the discussions of developers in Stack Exchange meta site \revision{which is the place encouraged by Stack Overflow for developers to raise their concerns about the various issues they observe in Stack Overflow.}
We qualitatively analyzed 1,697 posts in Stack Overflow where such discussions can happen. The posts contained discussions around the concerns related to reputation manipulation. 
Through an exploratory study of the post contents, we identified four
reputation fraud scenarios. The most common scenario is the formation of voting ring, i.e., \it{isolated} community of frauds who repeatedly 
upvote the posts of each other.To get revenge, users can unjustly downvote  the contents of good users, in an effort to intimidate those good users.

\nd\bf{$\bullet$ Study 2. Detection of suspicious reputations.} We sought to automatically detect suspicious reputations in Stack Overflow.
We developed two algorithms to detect reputation abusers in Stack Overflow. 
The design of the two algorithms was guided by a qualitative study of 74 posts from the \emph{meta.stackexchange} site, \revision{where developers discussed about their perception of how frauds occur and how fraudulent activities can be detected.} \revision{The first algorithm detects suspicious communities whose
members show a high level of interactions among themselves by posting similar contents, responding to each other’s questions in a very short time, and accepting each other’s answers. Once such a community is identified, our algorithm considers the community and users in the community as suspicious.} 
The second algorithm looks for a sudden jump in reputation scores of a user between two quarters. If such a jump is well beyond normal jump (e.g., much more than the average jump of active users), our algorithm considers the user as suspicious. We evaluated the two algorithms by downloading the entire history of reputation logs of the suspicious users reported by our algorithms. 
Stack Overflow online site provides a reputation history dashboard for each user. Stack Overflow can remove a particular reputation score, if the process used to gain the reputation was not appropriate (i.e., due to a suspicious activity). Each algorithm outputs a list of users having suspicious reputation. We evaluated each such user by investigating his personalized 
reputation history dashboard. 
We observed that around 60-80\% of those users got their reputation scores reduced by Stack Overflow due to suspicious activities. 

We find that while one or two users in a our detected suspicious community are penalized by Stack Overflow, 
not all the members in the community are routinely penalized. This means that those users not penalized by Stack Overflow remain undetected by Stack Overflow, and thus they may continue to participate in fraud scenarios. Therefore, Stack Overflow can leverage our proposed prone to fraud detection algorithm to improve their fraud detection scripts. In addition, despite getting their scores reduced multiple times, many suspicious users are still active in Stack Overflow. 
\revision{Our meeting with the stack overflow team confirmed this hypothesis. We presented suspicious cases detected by our algorithms to them and they verified and acknowledged that they were indeed true fraudulent cases. It means that our algorithms have the potential to further improve the fraud detection mechanisms of the platform. As a result, based on the stack overflow acknowledgment of our detected cases, we can assert that our proposed algorithm can complement stack overflow current detection mechanisms.} Therefore, research in software engineering should take note of the existence of these potential fraud users in Stack Overflow and the low quality or incorrect 
contents shared by the users during their fraudulent activities. Otherwise, the inclusion of such contents will create threats to any research that 
use Stack Overflow data, such as to create API documentation~\cite{Robillard-OndemandDeveloperDoc-ICSME2017}, to recommend answers to unanswered 
questions or specific programming needs in an IDE~\cite{Ponzanelli-PrompterRecommender-EMSE2014}, and so on.

In the following sections, we first offer the background by describing the Stack Overflow reputation mechanism in \sec\ref{sec:background} before 
reporting the results of our studies. We 
report the results of our study to learn about the reputation fraud scenarios in Stack Overflow in \sec\ref{sec:study-reputation-fraud-scenarios}. 
We describe the algorithms we developed to detect suspicious reputations and the evaluation of the algorithms in \sec\ref{sec:algo-voting-ring-detection}. 
\revision{We offer a discussion of the study results in \sec\ref{sec:discussion}} and the threats to validity of our work in \sec\ref{sec:threats}. 
We discuss the related work in \sec\ref{sec:related-work} and conclude in \sec\ref{sec:summary}. All the data that are used in this study are available at: \url{https://github.com/mazloomzadeh/Reputation-Gaming.git}

\section{Reputation System in Stack Overflow}\label{sec:background}

According to Stack Overflow~\cite{website:SO-ReputationSystem-2019}, 
\emt{Reputation is a rough measurement of how much the community trusts you; it is earned by convincing your peers that you 
know what you're talking about.}
Conceptually, the reputation of user $i$ 
at any given time $t$ is a function of three elements, `score', `status', and `role'. 
\begin{equation}
Reputation_t^i = \{Score_t^i, Status_t^i, Role_t^i\}
\end{equation}
\revision {The reputation gain occurs in the following situations:
\begin{inparaenum}
\item the question is voted up, \item the answer is voted up, \item the answer is accepted, \item accepting an answer, \item the edits get accepted, \item obtaining bounty, \item getting automatic bounty, and \item obtaining association bonus.
\end{inparaenum} } 
According to Stack Overflow~\cite{website:SO-ReputationSystem-2019}, 
\emt{The more reputation you earn, 
the more privileges you gain and the more tools you'll have access to on the site - at the highest privilege levels, 
you'll have access to many of the same tools available to the site moderators. That is intentional. We don't run this site; the community does!} 

In this paper, unless explicitly mentioned otherwise, we refer to `reputation score' of a user when we note `reputation' of a user.


\section{REPUTATION FRAUD SCENARIOS}\label{sec:study-reputation-fraud-scenarios}
In this section, we report the results of a qualitative study of 1,697 posts in Stack Overflow meta sites 
where developers reported different reputation fraud scenarios. The outcome is a 
catalog of four reputation fraud scenarios. We describe each scenario and present how they are reported by developers in Stack Overflow.

\subsection{Study Methodology}
Our study involves three major steps (see \fig\ref{fig:MethodologyOverall}):
\begin{enumerate}[leftmargin=10pt]
  \item \bf{Data Collection}: We collect posts where reputation fraud scenarios are discussed.
  \item \bf{Data Analysis}: We manually label reputation fraud scenarios discussed in the posts.
  \item \bf{Fraud Scenario Report}: We report the fraud scenarios by connecting those with specific examples from the posts.
\end{enumerate} We discuss the steps below.

\nd\bf{$\bullet$ Data Collection.} 
Stack Overflow does not share information about the reputation fraud scenarios. 
Therefore, in the absence of any specific guidelines or a particular dataset, we attempt to learn about the reputation fraud scenarios 
from the discussions of developers in two meta sites: meta.stackoverflow.com and meta.stackexchange.com. 
StackExchange is the parent site for the online developer forum Stack Overflow. Stack Exchange consists of a network 
of online Q\&A sites, including Stack Overflow. StackOverflow is the most actively-viewed site
in the network. Each StackExchange site contains a \it{meta} sub-site where
users consult rules/issues. We picked the two meta-sites because they
feature programming discussions and the two meta sites are the most popular sites among all meta sites in Stack Exchange network. 
\revision{Meta Stack Exchange\footnote{https://meta.stackexchange.com/help/whats-meta} is where users discuss Stack Overflow's operations, policies, and the Stack Exchange Q\&A network software. It facilitates communication among Stack Exchange users for discussions on website functionality, policies, and decisions reached by the community. Users of Stack Exchange can engage with Stack Overflow company by reporting bugs, recommending enhancements, or suggesting novel functionalities through Meta. Additionally, Meta provides a channel for Stack Overflow company, to engage with the community by seeking feedback on ideas, features, and network-wide policies.}
On March 9, 2018, we searched the sites using the query
`voting+ring'. We picked this search query, because in our informal analysis of Stack Overflow posts, we found that 
this phrase is commonly used to denote different
reputation frauds. The query returned 8,551 posts (4,265 from meta.stackexchange.com, 4,286 from meta.stackoverflow.com).
The posts contained discussions from 1,728
distinct users, 25 of them participated in both sites.
From the 8,551 posts, we randomly picked 1,697 posts. This
subset is a statistically significant representation of the 8,551 posts with a 99\% confidence interval.

\nd\bf{$\bullet$ Data Analysis.} We manually analyzed each post to determine whether the post contains discussion about a reputation 
fraud scenarios. If so, we give the type of fraud scenario a name by reading the contents of the post. 
To do the analysis, we use open card sorting~\cite{Hudson-CardSorting-InteractionDesignFoundation2013}. 
Specifically, our card sorting approach attempted to categorize each of the 1,697 posts as follows:
\begin{enumerate}[leftmargin=10pt]
\item\bf{Scenario.} The post discusses a reputation fraud scenario. If so, we give the scenario a name by analyzing the scenario type as discussed in the post contents.
\item\bf{Concerns.} The post discusses relevant information but not a reputation fraud scenario, e.g., developers' concerns.
\item\bf{Irrelevant.} The post contents are not related to any reputation fraud scenario.
\end{enumerate}

%
%

%
The categorization involved five steps between the second and third authors of this paper:
\begin{enumerate}[leftmargin=10pt]
  \item \bf{Coding Consultation.} The two authors consulted 97 posts together to prepare a coding guide.
  \item \bf{Reliability Analysis.} The authors then separately coded 58 additional posts.
  The percent agreement between them was 83.3\%, which was substantial (Cohen $\kappa=.632$).
  \item \bf{Coding Completion.} With substantial agreement level, we assumed that either of the authors can label other posts. The second author labeled other posts.
  \item \bf{Clustering.} The labels are revisited multiple times to finalize their coding and to group the scenarios into higher categories.
  \item \bf{Finalize Coding.} The two authors discussed together the overall coding and the list of final reputation fraud scenarios.
\end{enumerate} 
On September 2020, another qualitative analysis was performed by the first and fourth authors to find out if the categories have evolved or new categories have emerged.
\revision{Sometimes regular users may state or claim they have been the subject of fraud that turns out to be not true. When moderators are somehow involved, we feel the posts are more factual.  Thus, the first and fourth authors added `moderator' to the query `voting+ring' to find more quality posts. They searched the sites using the query `moderator+voting+ring' to see if any actions by moderators are also posted. The query found 130 posts (60 from meta.stackexchange.com and 70 from meta.stackoverflow.com) including question posts and answer posts. Both authors have  read the top 37 posts of meta.stackexchange and the top 30 posts returned by the query from meta.stackoverflow.com. For each question post, they read all its related answer posts, and for each answer post, they read its question post and all its other answer posts. Thus in total, we have read 343 posts.
We found the posts had just presented or showed the same fraud categories as obtained by `voting+ring' in the previous analysis.}
 This qualitative assessment of the posts published after March 2018 did not reveal a new category and/or evolution of identified suspicious behavior categories found based on 2018 search.  Interestingly, we found that very few new questions were posted and the majority of the posts published after 2018 were answers to question posts regarding `voting-ring' posted in 2018 or before.
\revision{The queries were only designed to show the possibility of finding suspicious activities through the qualitative analysis that was done.  For sure, more complicated queries would lead to more coverage and finding other forms of fraudulent activities. By running queries searching for `manipulation', we found 247 posts (131 Posts in meta.stakexchange.com, 116 posts in meta.stackoverflow.com) and for `reputation fraud' 272 posts (177 Posts in meta.stakexchange.com, 95 posts in meta.stackoverflow.com) in 2022.
We randomly picked 10 posts from the result of running new suggested queries on each meta site. In total, we analyzed 40 randomly selected posts. But we did not find any new gaming scenarios. We found the posts had just presented or showed the same fraud categories as obtained by `voting+ring' in the previous analysis.
We aimed to demonstrate the existence of gaming, not to exhaustively detect all types of frauds. The `voting+ring' query initially used is sufficient because by changing it we did not find any new gaming scenarios.}



\subsection{The Reputation Fraud Scenarios}\label{sec:results}
\begin{figure}[t]
\centering
	\centering
   	\includegraphics[scale=.55]{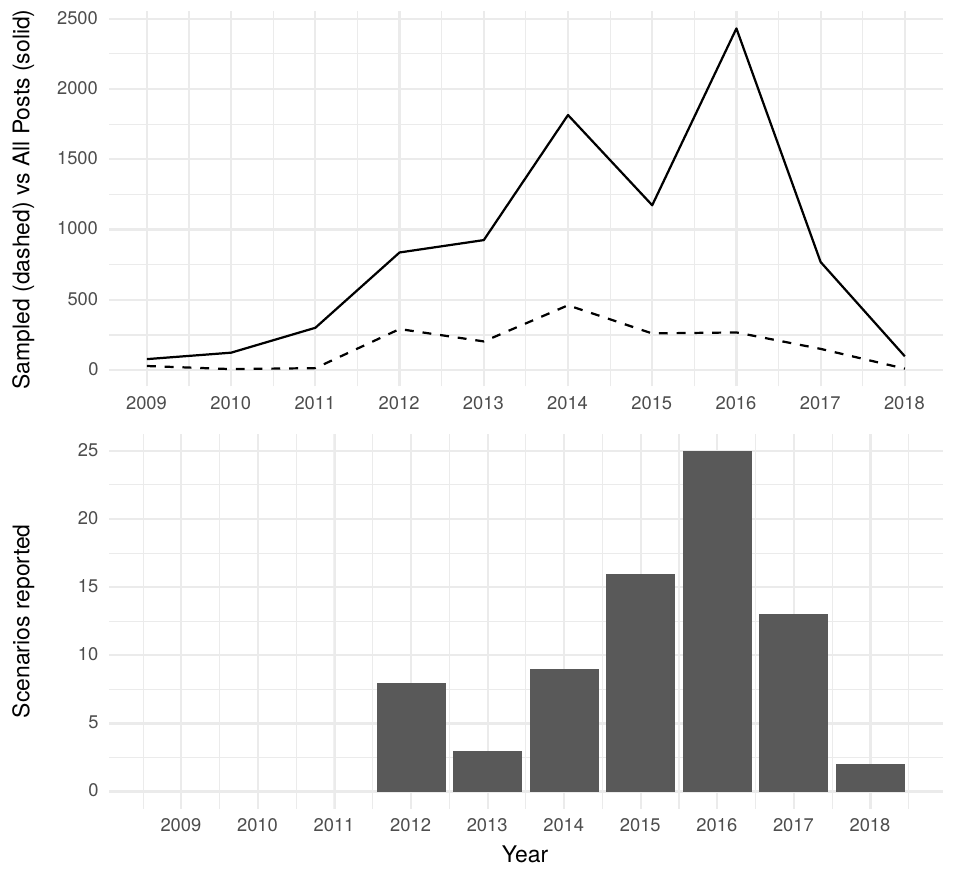}
   	\caption{The distribution of sampled (dashed) vs all posts (solid) and
   	the number of scenarios (bars) reported in the sampled posts by year}
   	 \label{fig:dataset}
\end{figure}

\begin{figure*}[tp]
\centering
	\centering

   \includegraphics[width=\textwidth,height=\textheight,keepaspectratio]{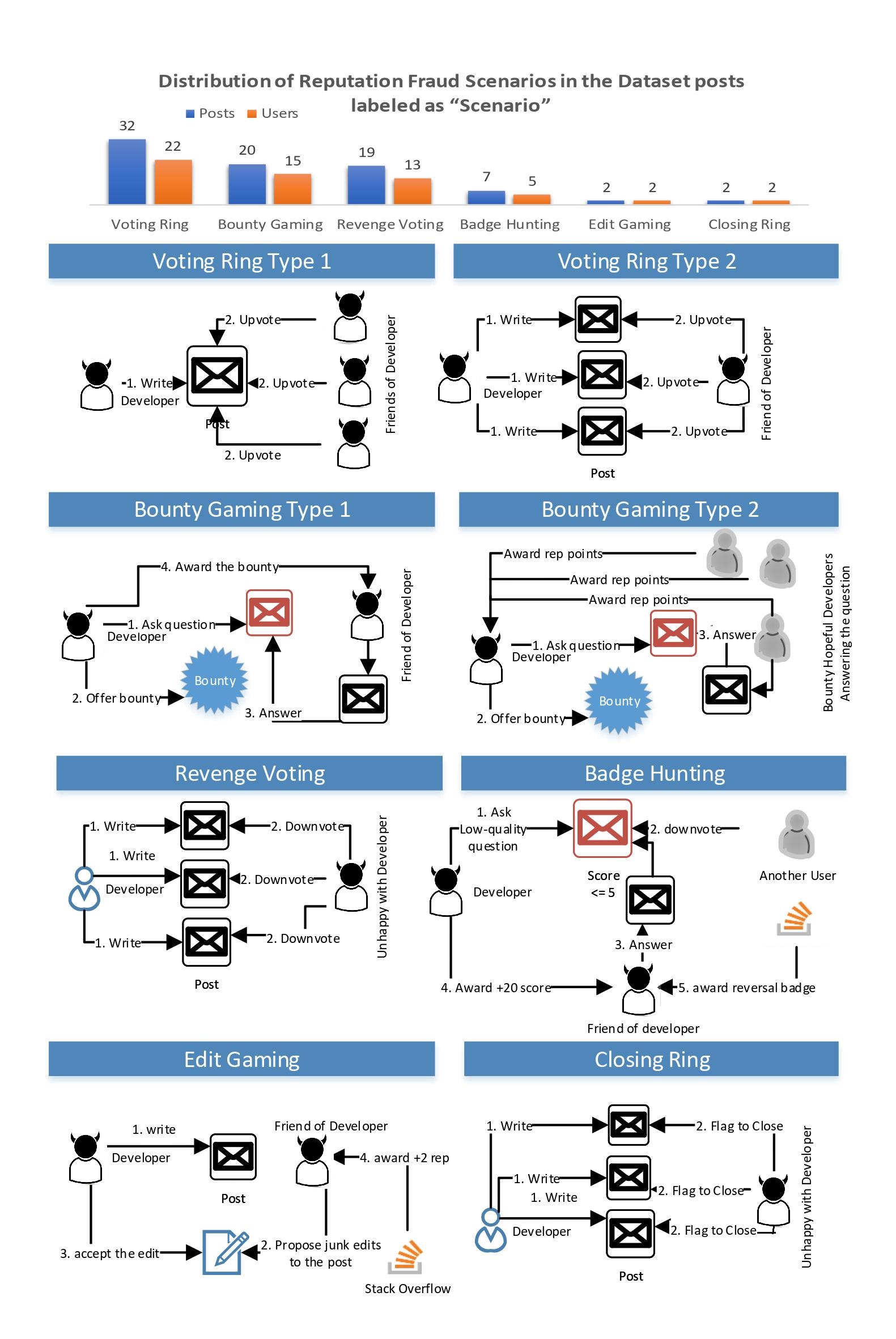}
   	\caption{The six types of reputation gaming scenarios as discussed in the 74 posts}
   	 \label{fig:wrongly-scenarios}
\end{figure*}

In \fig\ref{fig:dataset}, the top chart shows the distribution of all versus
sampled posts by years. The bottom histogram shows the scenarios reported by
years in the sample. The fraud scenarios are mentioned in 74
posts, solutions to those scenarios in 204 posts.
The other posts mainly explore the potential impact of the
scenarios.
We observed in total six types of reputation fraud scenarios.
The top bar-chart in \fig\ref{fig:wrongly-scenarios} shows their distribution. 
The voting ring scenario
is mentioned with the highest,
followed by bounty fraud. The
flowcharts illustrate the scenarios.
Each flowchart shows a sequence of events starting with a number and an
arrow denoting direction. We discuss each of the six scenarios in \fig\ref{fig:wrongly-scenarios}.

\subsection*
{Scenario {\number\value{scn}}. Voting Ring}
\stepcounter{scn}
A group of developers form a ring to upvote one another. We observed two types.

\subsubsection*{Type 1 Voting Ring} A developer asks a question and her
friends answer. Everyone in the ring upvotes
each other. Their posts can be trivial or even spams. 
\emt{Multiple StackExchange sites have been hit hard by what appears
to be a flood of students from a single university.
This has gotten real bad over the last few days on StackOverflow.
These students have created dozens of accounts that dump terrible questions,
have other accounts answer them using mostly plagiarized content, and then
vote in a tight ring to hugely inflate the votes on these posts
(\bf{MSO$_{A,322115}$}).}
The references $MSO_{m,n}$/$MSE_{m,n}$ denotes
the site (MSO/MSE), $m$ denotes the post type (Q = question, A =
Answer, C = Comment), $n$ denotes the post-id. MSO denotes
meta.stackoverflow.com and MSE meta.stackexchange.com.

\subsubsection*{Type 2 Voting Ring} A developer serially upvotes the
answers/questions of his friend. The posts can be part of multiple threads. In
Type 1, the ring-activity is limited to one thread at a time.
In Type 2, the activity spans multiple threads.

\subsection*{Scenario {\number\value{scn}}. Bounty Gaming}
In StackOverflow, a bounty is designed to promote an unanswered
question. To encourage
answers, the question asker offers 50-500 of his own
reputation to an answer that he will accept. We observed
two types of bounty gaming.
\stepcounter{scn}

\subsubsection*{Type 1 Bounty Gaming} Rather than accepting the best answer,
the asker can award it to
his friend who also answered.
\emt{I have recently encountered a very clear case of group bounty gaming.
question offered a +50 bounty.
I saw the bounty, answered it on the same day.
No response from the original poster for 5 days.
\ldots Today \ldots just mere hours before the bounty ended,
a new user popped up with an answer that is too similar to mine,
with mere changes in Class and some variable names. Within minutes, his answer
got accepted. (\bf{MSE$_{Q,141386}$})}

\subsubsection*{Type 2 Bounty Gaming} A developer places bounty to
gain upvotes to the question, rather than influencing others to answer. 
The developers in a ring
abuse this as follows: one developer offers the
bounty and other developers in the ring upvote him for placing the bounty.
\emt{I've seen a handful of questionable audit cases involving not particularly
great questions that had received a ton of upvotes.
In several of these cases, the voting could be traced back to a bounty
being placed on those questions. The reputation gained from votes
outweighs the 50 point bounty.(\bf{MSO$_{Q,300560}$)}
}


\subsection*{Scenario {\number\value{scn}}. Revenge Voting}
\stepcounter{scn}
A developer (victim) is targeted by one/more developers
(perpetrators). The victim's posts are downvoted by perpetrators to account
for a revenge. For example, the victim may have earlier labeled a post of a perpetrator as not
constructive (e.g., low quality).

\emt{I managed to get serial-downvoted.
2 times it was a massive downvote, 2 other times, it was under-the-radar. Both times,
the offender chose 3 or 4 of my top-rated questions and downvoted them.
(\bf{MSO$_{Q,356091}$})}

\subsection*{Scenario  {\number\value{scn}}. Badge Hunting}
\stepcounter{scn}
The \it{Reversal} badge is awarded to a developer who
provided a high-quality answer to a low-quality question. To get this badge, the
question must have scores $<=5$ and the answer
must get +20 scores. Developers in a ring can abuse when a developer asks a
question, another downvotes it to -5. Then another
developer answers the question and is awarded +20 scores.
\emt{I've seen a growing, frustrating trend, of people
answering egregiously bad, heavily-downvoted, off-topic, low-effort
questions. Then someone will upvote the answer \ldots then
another upvote \ldots and then "Here comes the reversal
badge!" (\bf{MSO$_{Q,277576}$})}

\subsection*{Scenario {\number\value{scn}}. Edit Gaming}
The developers can edit others' post. The post-owner
can accept/reject the edits. An accepted edit is awarded +2
score, a rejected edit is punished by -2 score. Reputation gamers can exploit
this by producing \it{meaningless} edits.
\stepcounter{scn}
\emt{Bad edits are a perennial topic here. The problem seems to be getting worse.
At the +2 rep per edit
rate, many users are simply making small and meaningless changes for the reputation.}\bf{MSO$_{Q,267311}$}

\subsection*{Scenario {\number\value{scn}}. Closing Ring}
\stepcounter{scn}
The posts of a developer can be intentionally flagged for
closure by other developers. We observed that developers suspected the existence of fraudulent \it{closing ring}.
\emt{"Closing rings" are an interesting beast. We know they exist, but the jury's still out on whether or not they're inherently malicious or how we'd go about
investigating them.} \bf{MSE$_{A, 161870}$}. \\ 

\begin{figure*}[tp]
\centering
	\centering

   \includegraphics[width=\textwidth,height=\textheight,keepaspectratio]{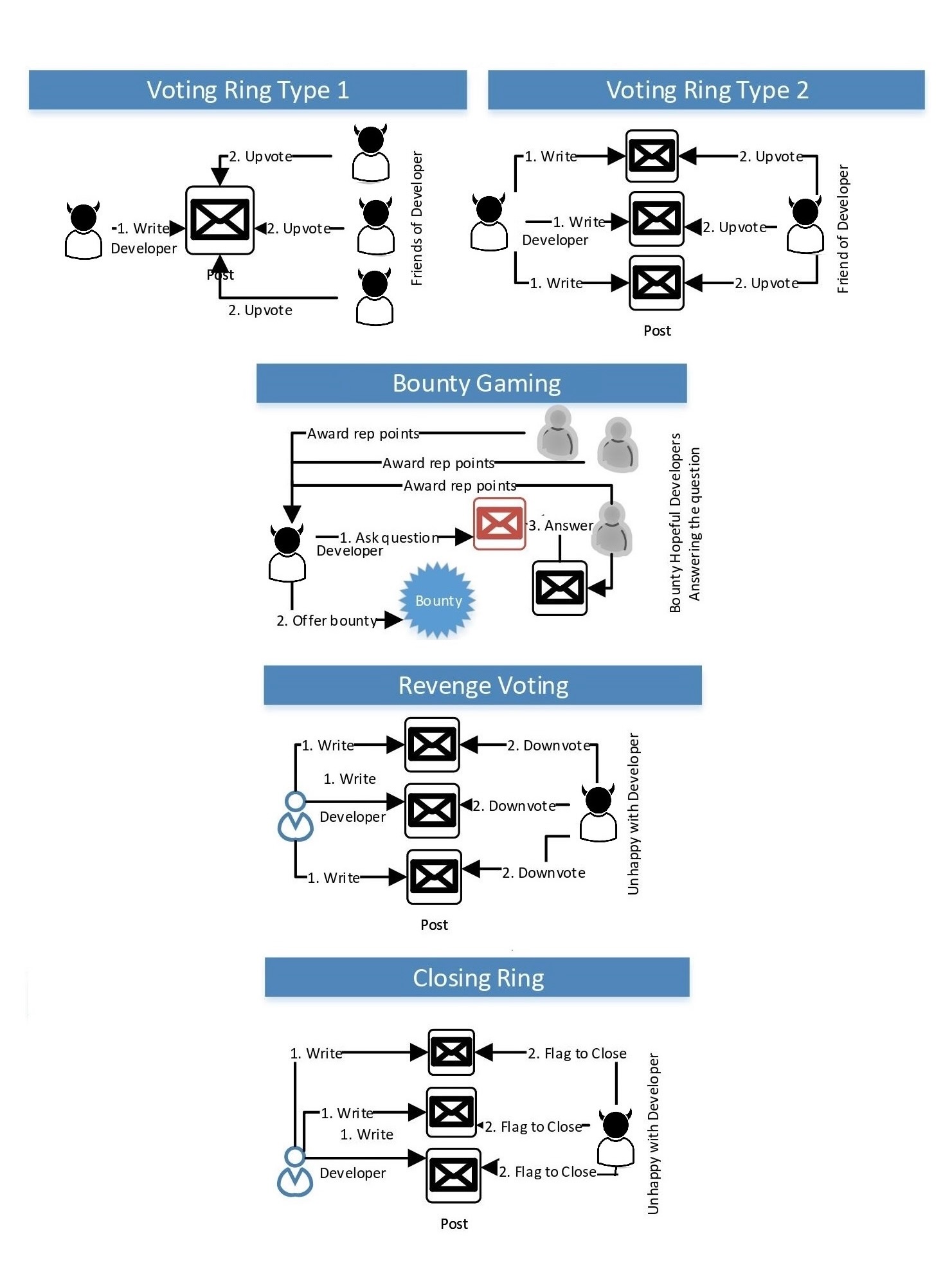}
   	\caption{The final four types of reputation gaming scenarios acknowledged by the Stack Overflow team}
   	 \label{fig:scenarios}
\end{figure*}

\revision{For assessing our six fraud scenarios, we contacted several StackOverflow moderators and regular users.  Based on their recommendations, we checked the reputations of people we had used for referring to our findings as evidence.  Additionally, we also checked whether the found scenarios are still valid in SO. Interestingly, the majority of evidence that we had collected throughout our study was stated by moderators. Based on a thorough analysis of scenarios, we took three scenarios out of our suspicious categories:
\begin{inparaenum}[(1)]
\item Type 1 Bounty Gaming was omitted since the poster had a low reputation, \item Edit gaming was removed as one of SO members mentioned, any form of gaming by abusing the gains from edit points does not make a large impact on reputation, \item Badge Hunting was removed; although a moderator pointed to a fraud formed by the Reversal badge, the badge is now expired.
\end{inparaenum} As a result, we presented the four final fraud scenarios acknowledged by SO team in Figure 4. In the following, you can find the description of the four final fraud scenarios.}

\newcounter{var1}
\stepcounter{var1}
\subsection*{\revision{Scenario {\number\value{var1}}. Voting Ring}
\stepcounter{var1}}
\revision{A group of developers form a ring to upvote one another. We observed two types.}

\subsubsection*{\revision{Type 1 Voting Ring}} \revision{A developer asks a question and her friends answer it. As shown in \fig\ref{fig:scenarios} everyone in the ring upvotes each other. Their posts can be trivial or even spam. \acceptance {A moderator once said}: \emt{Multiple StackExchange sites have been hit hard by what appears to be a flood of students from a single university. This has gotten real bad over the last few days on StackOverflow. These students have created dozens of accounts that dump terrible questions, have other accounts answer them using mostly plagiarized content, and then vote in a tight ring to hugely inflate the votes on these posts.} (\bf{MSO$_{A,322115}$}). The references $MSO_{m,n}$/$MSE_{m,n}$ denotes the site (MSO/MSE), $m$ denotes the post type (Q = question, A = Answer, C = Comment), $n$ denotes the post-id. MSO denotes meta.stackoverflow.com and MSE denotes meta.stackexchange.com.}

\subsubsection*{\revision{Type 2 Voting Ring}}\revision{ A developer serially upvotes the
answers/questions of his friend. The posts can be part of multiple threads. In Type 1, the ring-activity is limited to one thread at a time.
In Type 2, the activity spans multiple threads.}

\subsection*{\revision{Scenario {\number\value{var1}}. Bounty Gaming}
\stepcounter{var1}}
\revision{In StackOverflow, a bounty is designed to promote an unanswered
question. To encourage
answers, the question asker offers 50-500 of his
reputation to an answer that he will accept.
A developer places bounty to gain upvotes to the question, rather than influencing others to answer. 
The developers in a ring
abuse this as follows: one developer offers the
bounty and other developers in the ring upvote him for placing the bounty. \acceptance{A moderator said}:
\emt{I've seen a handful of questionable audit cases involving not particularly
great questions that had received a ton of upvotes.
In several of these cases, the voting could be traced back to a bounty
being placed on those questions. The reputation gained from votes
outweighs the 50 point bounty.} (\bf{MSO$_{Q,300560}$}).}


\subsection*{\revision{Scenario {\number\value{var1}}. Revenge Voting}
\stepcounter{var1}}
\revision{In this form of gaming, a developer (victim) is targeted by one/more developers (perpetrators). The victim's posts are downvoted by perpetrators to account for revenge. If the perpetrator downvotes an answer, he loses one reputation per downvote.  However, if he downvotes questions, he does not lose any reputation score. For example, the victim may have earlier labeled a post of a perpetrator as not constructive (e.g., low quality).  In this scenario, the downvoter loses 2 points. \acceptance{A moderator said}: \emt{I managed to get serial-downvoted. 2 times it was a massive downvote, 2 other times, it was under-the-radar. Both times, the offender chose 3 or 4 of my top-rated questions and downvoted them.} (\bf{MSO$_{Q,356091}$}).}

\subsection*{\revision{Scenario {\number\value{var1}}. Closing Ring}
\stepcounter{var1}}
\revision{The posts of a developer can be intentionally flagged for
closure by other developers. We observed that developers suspected the existence of fraudulent \it{closing ring}. \acceptance{A moderator and staff said}: 
\emt{"Closing rings" are an interesting beast. We know they exist, but the jury's still out on whether or not they're inherently malicious or how we'd go about
investigating them.} (\bf{MSE$_{A, 161870}$}).}

\begin{tcolorbox}[opacityback=0, standard jigsaw, title=Summary of Study 1 \hrule
\it{Understanding Reputation Gaming Scenarios}]
We manually categorized 1,697 posts from Stack Overflow meta sites to understand the types of reputation manipulation scenarios that users discussed in those posts. We observed four distinct reputation gaming scenarios: \begin{inparaenum}[(1)]
\item Voting Ring,
\item Bounty Gaming, 
\item Revenge Voting, and 
\item Closing Ring.
\end{inparaenum} The `Revenge Voting' and `Closing Ring' scenarios are used to purposefully reduce the reputation scores of a user. 
The other scenarios are used to illegally increase the reputation scores. The voting ring scenario is the most commonly discussed, i.e., it is the most prevalent reputation gaming scenario 
in Stack Overflow. 
\end{tcolorbox}

\section{Suspicious Reputations}\label{sec:algo-voting-ring-detection}
In this section, we describe the design and development of algorithms to automatically detect suspicious reputations in Stack Overflow. 
We specifically focused on the frauds that attempt to exploit the Stack Overflow voting system to increase their reputation scores. 
As we noted before in \sec\ref{sec:study-reputation-fraud-scenarios}, two (Voting Ring, Bounty Gaming) out of the four gaming scenarios  are related to increase of reputation scores. 
This is because, the purpose of reputation gaming in Stack Overflow is to gain reputation scores. Out of the two scenarios the most common gaming scenario was voting ring, where users form a ring (i.e., a community) 
to upvote the posts of each other. Our proposed two algorithms are designed and evaluated to detect users who are either part of a suspicious community 
or who show unusual jump in their reputation scores in a short time.
\subsection{Study Methodology}
Stack Overflow does not share information about reputation frauds. Stack Overflow also removes 
several fields in its data dumps that are needed to detect reputation frauds, the tracking of upvotes and downvotes to a post (e.g., who downvoted a post and when the downvote happened). Stack Overflow may remove suspicious posts in Stack Overflow using its internally developed scripts. However, we observed that while Stack Overflow removes 
posts that are suspicious and removes the corresponding scores from a suspicious user, the user record is not deleted. Therefore, we can develop algorithms to find such 
suspicious reputations in Stack Overflow. To detect the suspicious reputations, we performed the following three major steps  (see \fig\ref{fig:MethodologyOverall}).
\begin{enumerate}[leftmargin=10pt]
    \item \bf{Learn Reputation Fraud Characteristics (\sec\ref{subsec:learning-indicators-reputation-fruad})}.
        We analyze Stack exchange posts related to fraud
    discussions to understand the specific indicators of reputation frauds.

    \item \bf{Develop Algorithms to Detect Suspicious Reputations (\secs\ref{subsec:algo1-gaming-community}, \ref{subsec:algo2-repgamer}).}
    We design two algorithms based on the reputation gaming scenarios. The inputs to an algorithm are Stack Overflow posts and users who participated in the posts. The output is a list of potential suspicious reputations.

	\item \bf{Evaluation (\sec\ref{subsec:evaluation-algo-reputation-suspicious}).} We evaluate the two algorithms by analyzing the outputs, i.e., the list of suspicious users. For each such user in our evaluation corpus, we check whether the reputation score of the user was reduced by Stack Overflow due to fraud/non-recommended activity.
\end{enumerate}



\subsection{Reputation Fraud Indicators}\label{subsec:learning-indicators-reputation-fruad}
With a view to design algorithms to detect users in Stack Overflow who abuse the reputation system to increase their scores, 
we first need to understand the characteristics of the frauds during their exploitation. In the absence of 
specific guidelines of reputation fraud indicators, we attempted to learn such indicators from the discussions of Stack Overflow 
users similar to our previous study in \sec\ref{sec:study-reputation-fraud-scenarios}. Our analysis consisted of 
all the 26 posts related to voting ring discussions. In addition, we also analyzed 48 additional posts 
from meta.stackexchange.com where voting rings are discussed. In total, we analyzed 74 posts. After the 74 posts, no new indicators or themes emerged. 

The first two authors analyzed the post contents together and identified the indicators of reputation frauds that were discussed in the posts. 
The process consisted of around three weeks of discussions and multiple collaborative sessions between the two authors over email and Skype. 
We followed the open card sorting process for each post, where both authors discussed together the specific indicator that was discussed in the post. 
As the authors discussed the indicators, themes and patterns emerged. At the end, we observed two types of indicators:
\begin{enumerate}[leftmargin=10pt]
  \item \bf{Suspicious Community.} This indicator points to a group of users who form a community to increase their reputations in various ways. 
  \item \bf{Suspicious User.} This indicator points to individual users who may not exhibit affiliation to a specific community, but nevertheless attempt to exploit the 
  reputation mechanism by quickly gaining scores through diverse tactics. 
\end{enumerate}  
\begin{figure}[t]
\centering
	\centering
	\hspace*{-.7cm}%
   	\includegraphics[scale=.8]{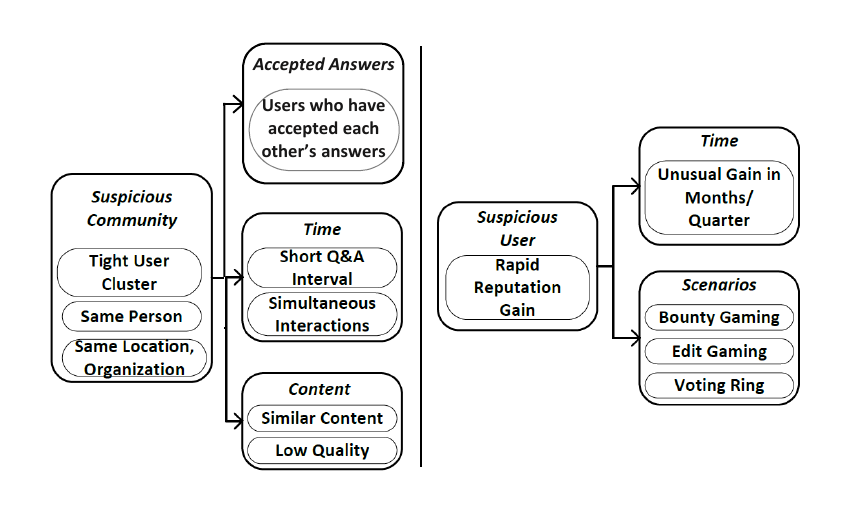}
   	\caption{The major indicators of reputation frauds found in our study}
   	 \label{fig:Replacement_for_Figure4_CharacterizeReputationFraud}
\end{figure} In \fig\ref{fig:Replacement_for_Figure4_CharacterizeReputationFraud}, we show two flowcharts to illustrate how the indicators are leveraged to help find out potential  reputation frauds. We discuss the flowcharts below.

\nd\bf{$\bullet$ Suspicious Community.} A group of friends can form a closely knit community to upvote the posts of each other. While 
in Stack Overflow, a user cannot prevent another user from responding to a question, the frauds in a community can prevent other users 
by using two tactics: \begin{inparaenum}[(1)]
\item A user in a community has asked a question and another user instantly replied to the question. The first user then accepts the answer. The two users then upvote the question 
and answers. 
\item The users post multiple similar questions, allowing other users in the community to post similar answers.  
\end{inparaenum}  According to discussions around this indicator in meta.stackexchange.com: 
\emt{A number of new users post, answer, and accept each others' off-topic questions before we can close them (in less than 5 min). Accepting each others' answers gains them enough reputation to start up-vote each others' posts.}
In a suspicious community, the same person may also have multiple user ids. This can happen when a user can open multiple accounts in Stack Overflow, 
thereby allowing one account posting questions and another account to answer to the question. Such a community can also be formed by 
users residing in a similar physical location or organization. According to discussions around this indicator in meta.stackexchange.com, we found an evidence: 
\emt{As answered in the other question it seems the two accounts of you were that tightly related (only answering and voting on each other) that a moderator assumed the two of you have a voting ring.
Since serial voting (or voting on a specific user only) is not allowed, the moderator deleted the other user's account and suspended yours.
The lesson is: don't just vote on one user. Also try to be open to other users. Be careful when sharing your posts with other users on social media (\bf{MSE$_{A,289558}$}).} 

\nd\bf{$\bullet$ Suspicious User.} While frauds can work in a tight cluster, to evade detection by Stack Overflow, some frauds may also try to exploit without showing obvious connection to 
one or a group of users. Therefore, such users can be harder to detect. Such frauds typically exhibit unusual gain in their reputation scores in a short time. 
A stackoverflow user was surprised at the gain of another user in a discussion:
\emt{It's been 14 days you created your profile, but you don't sound new to this site. 
Because you got 14 Badges already. I am very much impressed. Yesterday you post your first flag to moderator attention. 
Your age is 20 so believe that you are a college person. So your upvotes may be a prank of your friends (but I am not sure for it). 
In 14 days you have done 192 activities and your profile view is 110? Your profile views says that someone is visiting your profile frequently for those false upvote (\bf{MSE$_{A,129086}$}).}
As stated before, according to discussions in meta.stackexchange.com, another evidence is:
\emt{I deleted your account on November 21. You and several others all clearly conspired to defraud the voting system by participating in a voting ring that was propping up low-quality or incorrect content. This voting coordination came to the attention of moderators through community flags or through our standard tools.
As your account had contributed little of value beyond the coordinated voting, I felt it safe to delete it in order to invalidate these votes. The friends of yours with more significant contributions were strongly warned about this and their votes manually invalidated by Stack Exchange employees.
In the future, I highly recommend not coordinating votes between yourself and your friends in an effort to gain an unfair advantage over others (\bf{MSE$_{A,272729}$}).}
\subsection{Algorithm 1. Suspicious Community Detection}\label{subsec:algo1-gaming-community}
Based on indicators of suspicious community that we learned in \sec\ref{subsec:learning-indicators-reputation-fruad}, we design an algorithm to automatically detect suspicious reputations that 
collaborate by forming communities to increase their reputation scores; abusing Stack Overflow reputation mechanism. 
Given that sensitive user information is hidden by Stack Overflow (see \sec\ref{subsec:learning-indicators-reputation-fruad}), 
as we noted in \sec\ref{subsec:learning-indicators-reputation-fruad}, we 
developed an algorithm that utilizes publicly available data in Stack Overflow (i.e., without the sensitive data) based on the findings 
we discussed in \sec\ref{subsec:learning-indicators-reputation-fruad}.
In our analysis, we only considered edges among users who have answered at least one of each other's questions in a round-robin fashion. Other forms of questioning and answering are not used to create edges among users since those forms of communication were not considered suspicious based on our assumptions.  That is if one user only answered a question asked by another user who had not had any other interactions, we did not consider them as related. Definitely, more possible forms of gaming based on different assumptions may exist that require other forms of users’ communities.  However, we have not considered them in this study.As a result,
our algorithm relies on the following assumption; \it{we selected every two users who have answered at least one of each other's questions. The result of this selection was 21,072 users, 46,822 question posts, and 50,331 answer posts.}
As the left flowchart in \fig\ref{fig:Replacement_for_Figure4_CharacterizeReputationFraud} shows, this group of users can become suspicious when 
\begin{inparaenum}[(1)]
  \item  The group of users is only found to be asking and answering to each other, accepting answers of each other and preventing other users' answers from being accepted, i.e., an isolated community which has a high number of interaction links, every one of which shows an answer given by one of the community's users to a question asked by another user of that community and has been accepted by the questioner.   
  \item The group of users are using similar contents in multiple questions and answers among themselves.
  \item The group of users are communicating in a very short time interval, i.e., an isolated community whose users are asking questions and answering each other during a very short time interval and the community has a lot of interaction links, every one of which shows an answer given by one of the community's users to a question asked by another user of that community.
\end{inparaenum} 

The flexibility of the above heuristics allows us to implement those using publicly available Stack Overflow data. For example,  \begin{inparaenum}[(1)]
\item we can use the post meta data to learn about ids of the users that asked questions or answered to questions 
\item we can use post creation time to learn about the time of the posting, and the id of the accepted answer to a question.
\item We can use post contents to determine the similarity of contents between posts.
\end{inparaenum} Our algorithm consists of the following steps:
\begin{enumerate}[leftmargin=10pt]
  \item Create an \bf{Interaction Table} by selecting every two users who have answered at least one of each other's questions and focused on their posts. 
  \item Create an \bf{Interaction Graph} out of the Interaction Table by connecting the askers and answerers in {Interaction Table}.   
  \item \bf{Detect Community} in the {Interaction Graph}. 
  \item \bf{Detect Suspicious Community $GC_{V1}$} as an isolated community whose members respond to the posts of each other in a very short time and the community has a lot of links, every one of which shows an answer given by one of the community's member to a question asked by another member of that community.
  \item \bf{Detect Suspicious Community $GC_{V2}$} as an isolated community that has a lot of links, every one of which shows an answer given by one of the community's users to a question asked by another user of that community and has been accepted by the questioner.
  \item \bf{Detect Suspicious Community $GC_{V3}$} as a community that is isolated and the questions they ask and the answers they post are mostly similar to each other.
\end{enumerate} We describe the steps below.

\begin{inparaenum}
\item\bf{Create Interaction Table.} We take as input the Posts.xml from Stack Overflow and create a table with following columns: \begin{inparaenum}[(i)]
\item QuestionId, 
\item AskerId,
\item AnswerId,
\item AnswererId,
\item QuestionCreationTime,
\item AnswerCreationTime,
\item IsAcceptedAnswer,
\item QuestionBody, and
\item AnswerBody
\end{inparaenum} 

\item\bf{Generate Interaction Graph.} We take as input the Interaction Table and generate a graph of users as follows. Each node in the graph is a unique user id (AskerId/AnswererId). An edge between two nodes (an AskerId and an AnswererId) is established when one user answers a question asked by another user (this edge is from AskerId to AnswererId). Each edge between two nodes $i$ and $j$ is given a vector of weight as: 
\begin{equation}
E_{i,j} = \{E_{key} , E_{accepted} , E_{time}\}
\end{equation} $E_{key}$ which is a unique token like "QuestionId/AnswerId" denotes the key of the edge. $E_{accepted}$ denotes whether a given answer to a question has been accepted or not. $E_{accepted}$ is 1 if user $i$ has accepted the answer given by user $j$ and 0 otherwise. $E_{time}$ is the time that took for $j$ to answer a question asked by user $i$. Also, we define $W(i, j)_{total\textunderscore edges}$ which denotes the frequency of \textit{complete} interactions between i and j and whose value equals to $W(j, i)_{total\textunderscore edges}$; so it does not matter we write $W(i, j)_{total\textunderscore edges}$ or $W(j, i)_{total\textunderscore edges}$.
For example, if user denoted by node $i$ has asked 2 questions that user denoted by node $j$ has answered and if $j$ has asked 1 question that $i$ has answered, then $W(i, j)_{total\textunderscore edges} = 3$.

\item\bf{Detect Community.} We take as input the Interaction Graph and detect community of users in the graph. A community is identified when users in a group show  
more interactions among them 
than the rest of users in Stack Overflow. This means that the problem for us to solve is 
to partition our Interaction Graph into communities of densely connected nodes, with the nodes belonging to different 
communities being only sparsely connected. Therefore, we need a community detection algorithm that is good at finding such mostly isolated 
communities, while at the same time being able to operate on a large graph with millions of nodes (such as ours). The Louvain algorithm~\cite{Blondel-FastUnfoldingCommunitiesLargeNetworks-JournalStat2008} 
satisfies both requirements. The Louvain algorithm detects communities in a graph by partitioning the graph into sub-networks ~\cite{Blondel-FastUnfoldingCommunitiesLargeNetworks-JournalStat2008} based on the definition of a metric `Modularity' ~\cite{newman2004analysis}. 
Modularity measures the density of links (i.e., edges) between nodes inside a community as opposed to the density of links of the community with other communities. 
The metric $Q$ ranges between -1 and 1 and is calculated as:   
\begin{equation}
Q = \frac{1}{2m}\sum_{i,j}\bigl[ A_{ij} - \frac{k_ik_j}{2m}\bigr]\delta (c_i, c_j)
\end{equation} Here $A_{ij}$ is the weight between nodes $i$ and $j$. For our problem, the weight we chose is $W(i, j)_{total\textunderscore edges}$, i.e., the total number of edges between two users $i$ and $j$. 
$k_i = \sum_j A_{ij}$ is the total link weight attached to node $i$ and $k_j$ is likewise for node $j$. $m$ is equal to \(\frac{1}{2}{\sum_{i,j} A_{ij}}\). Thus $\bigl[ A_{i,j} - \frac{k_ik_j}{2m}\bigr]$ 
measures how strongly nodes $i$ and $j$ are connected to each other, compared to how strongly they could have been connected in a random network (i.e., without any real connection). 
$\delta$ is called 
the Kronecker delta, it is 1 when nodes $i$ and $j$ are assigned to the same community and 0 otherwise. In order to determine whether two nodes $i$ and $j$ belong to a community, the 
modality value between those needs to be optimal. The Louvain method finds the optimal value by first assigning each node to its individual community and then 
repeatedly assigning and removing other nodes into a community, until the modularity value is observed as the optimal. The change in modularity is calculated 
as follows.
\begin{equation}
\Delta Q = \bigl[\frac{\sum_{in}+k_{i,in}}{2m} - \bigl(\frac{\sum_{tot}+k_i}{2m} \bigr)^2\bigr] - \bigl[ \frac{\sum_{in}}{2m} - \bigl( \frac{\sum_{tot}}{2m}\bigr)^2 -\bigl( \frac{k_i}{2m}\bigr)^2 \bigr]
\end{equation} 

\item\bf{Detect Suspicious Community $GC_{V1}$.} Given as input all the communities returned by Louvain algorithm on our Interaction Graph, this step heuristically labels some of the communities as suspicious based on three heuristics: 
\begin{inparaenum}[\bfseries (H1)]
\item the community is isolated, 
\item the number of all edges (answers given by the community's users to questions asked by other users of that community) in that isolated community is above a given threshold $\tau_{L}$.The higher the number, the more likely the users in the community are suspicious. And 
\item The members of the community answer a question asked by another member of the community in a very short time $\tau_{T}$, which means the value of feature $E_{time}$ of all links in that community is a very short time (under or equal threshold $\tau_{T}$).
\end{inparaenum} 
\begin{figure}[t]
\centering
	\centering
	\hspace*{-.7cm}%
   	\includegraphics[width=\textwidth,height=\textheight,keepaspectratio]{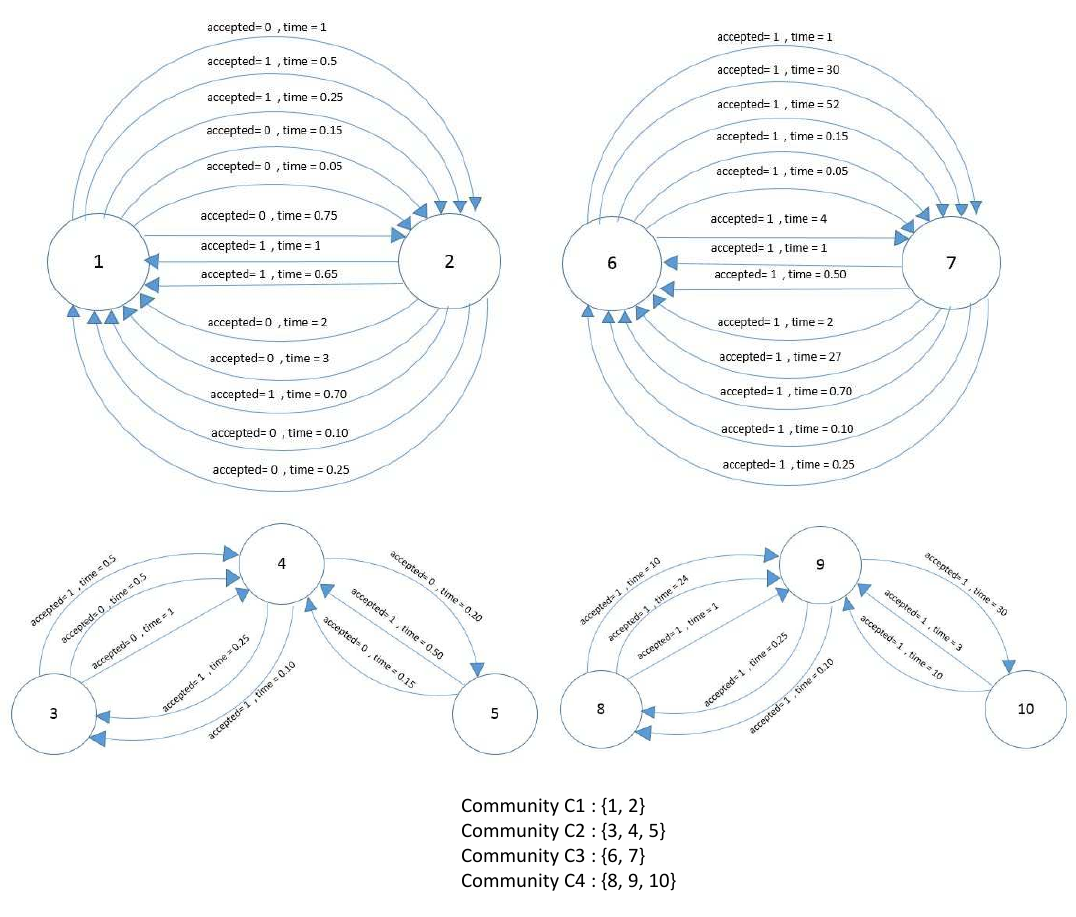}
   	\vspace{-5mm}
   	\caption{The formation of isolated communities}
   	 \label{fig:Replacement_for_Figure5_IllustrationSuspiciousCommunity}
\end{figure} As an illustration, consider the communities $C_1$ and $C_2$ shown in \fig\ref{fig:Replacement_for_Figure5_IllustrationSuspiciousCommunity}. The first community detected is $C_1$ with two nodes (1, 2). 
An edge from user 1 to user 2 shows that user 2 has answered one of the question of user 1. Also, each edge has a key which is "QuestionId/AnswerId" which we have not shown in \fig\ref{fig:Replacement_for_Figure5_IllustrationSuspiciousCommunity} for more readability and has two features $E_{accepted}$ and $E_{time}$; for example, an edge from user 1 to user 2 has two features $E_{accepted}$ which would be set to 1 if user 1 has accepted the answer of user 2 and 0 otherwise and $E_{time}$ which shows the time took by user 2 to answer the question of user 1. In our case, the total number of edges (answers given by the community's users to questions asked by other users of that community) between nodes (1, 2) is 13. All members in $C_1$ answer the questions of each other in less than 24 hours. The second community $C_2$ consists of three nodes (3, 4, 5). The members of $C_2$ have answered the questions of each other in 24 hours as well. 
Both of the two communities are isolated. Therefore, if our threshold for $\tau_{T} = 24H$ and for $\tau_{L} = 8$ , then both of these communities are labeled as suspicious by our algorithm $GC_{V1}$.

\item\bf{Detect Suspicious Community $GC_{V2}$.} Given as input all the communities returned by Louvain algorithm on our Interaction Graph, this step heuristically labels some of the communities as suspicious based on three heuristics: 
\begin{inparaenum}[\bfseries (H1)]
\item The community is isolated, i.e., the out degree from a given community is 0. 
\item The number of all edges (answers given by the community's users to questions asked by other users of that community) in that isolated community is greater equal to a given threshold $\tau_{L}$. The higher the number, the more likely the users in the community are suspicious.
\item all users of the community have accepted all the answers given by the other users of that community to their questions, i.e., the value of feature $E_{accepted}$ of all edges in that community is 1.
\end{inparaenum}

As an illustration, consider the communities $C_3$ and $C_4$ shown in \fig\ref{fig:Replacement_for_Figure5_IllustrationSuspiciousCommunity}. The third detected isolated community is $C_3$ with two nodes (6, 7) and 13 edges, which shows that these two users have answered each other's questions 13 times. The value of feature $E_{accepted}$  of all edges on this community is 1 which means that all users of the community have accepted all the answers given by the other users of that community to their questions. For example, user 7 has accepted all the answers given by user 6 to his/her questions and user 6 has accepted all the answers given by user 7 to his/her questions. The fourth detected isolated community is $C_4$ with three nodes (8, 9, 10) and 8 edges, which shows these three users have answered each other's questions 8 times. The community's users have accepted all of each other's answers.  Therefore, if our threshold for $\tau_{L} = 8$, then both communities $C_3$ and $C_4$ are labeled as suspicious by our algorithm $GC_{V2}$. Thus, with our three heuristics (H1, H2, H3), we are more confident of community $C_3$ being suspicious.

\begin{figure*}[tp]
\centering
	\centering

   \includegraphics[width=0.7\textwidth,height=\textheight,keepaspectratio]{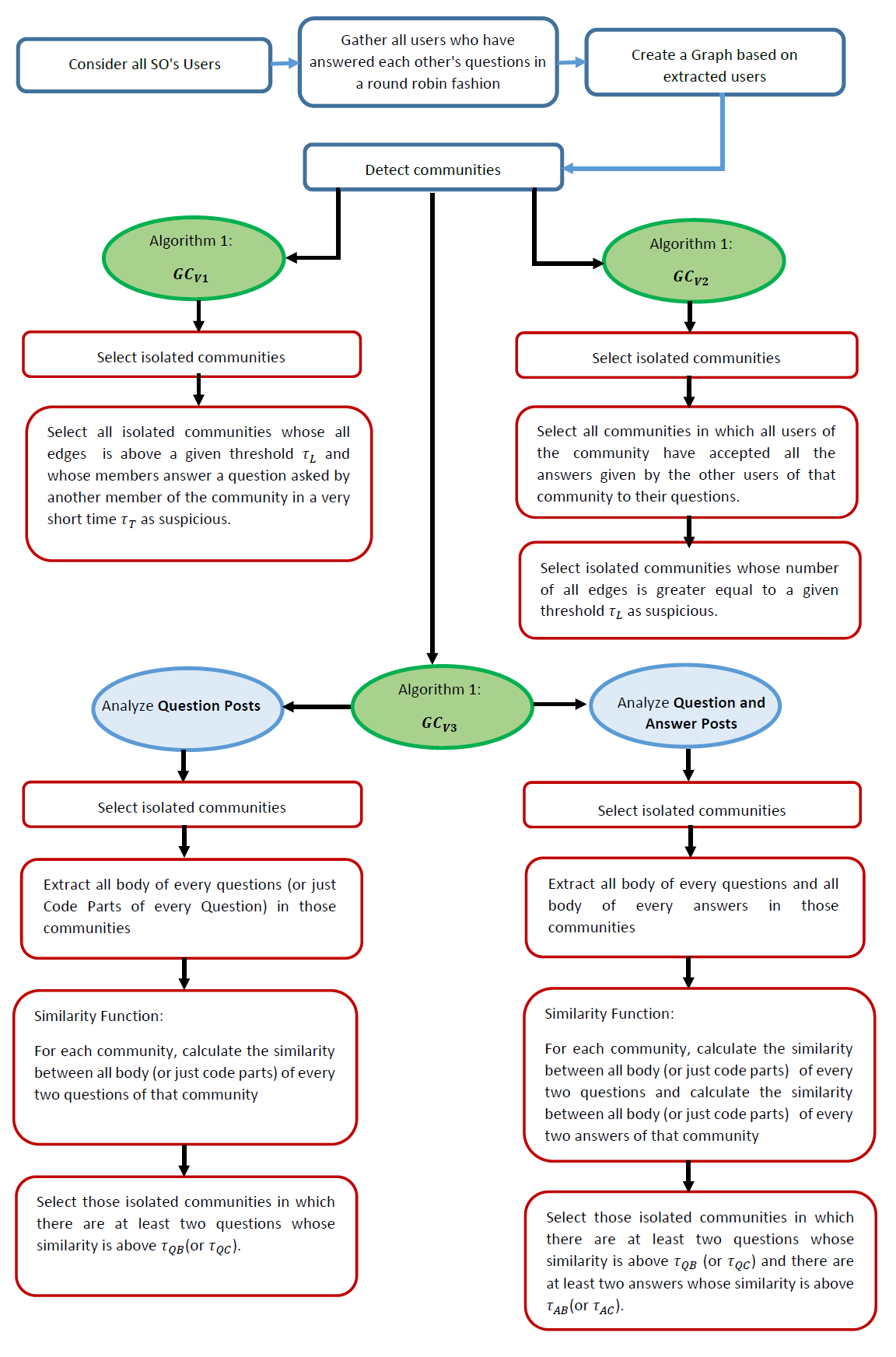}
   	\caption{The detailed description of algorithm 1}
   	 \label{fig:algorithm1-detailed-description}
\end{figure*}

\begin{figure}[h]
\centering
\begin{minipage}{0.4\textwidth}%
\centering
\subfloat[Subfigure 1 list of figures text][]{
\includegraphics[width=0.59\textwidth]{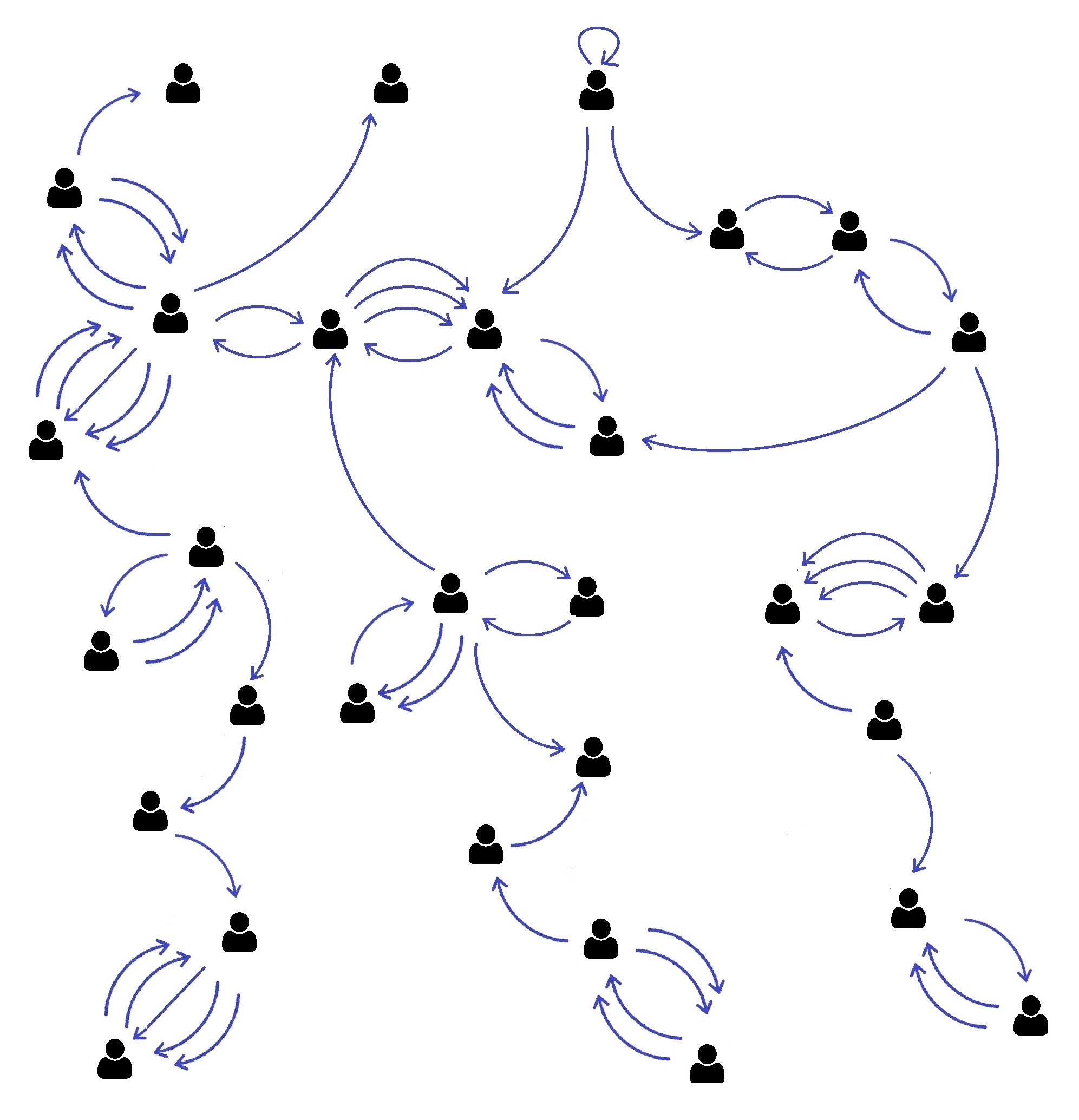}
\label{fig:subfig1}}\\
\subfloat[Subfigure 2 list of figures text][]{
\includegraphics[width=0.59\textwidth]{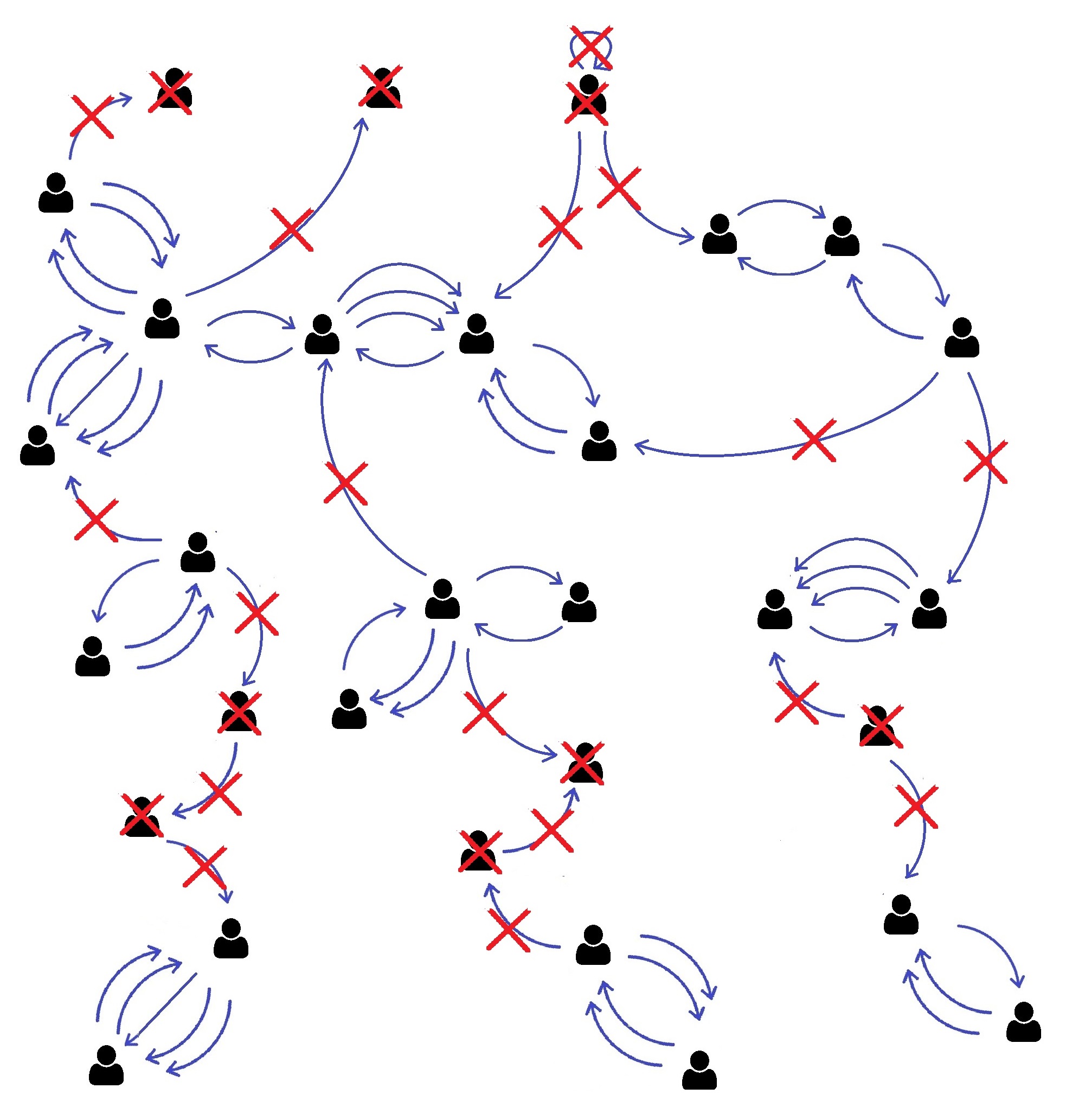}
\label{fig:subfig2}}\\
\subfloat[Subfigure 3 list of figures text][]{
\includegraphics[width=0.59\textwidth]{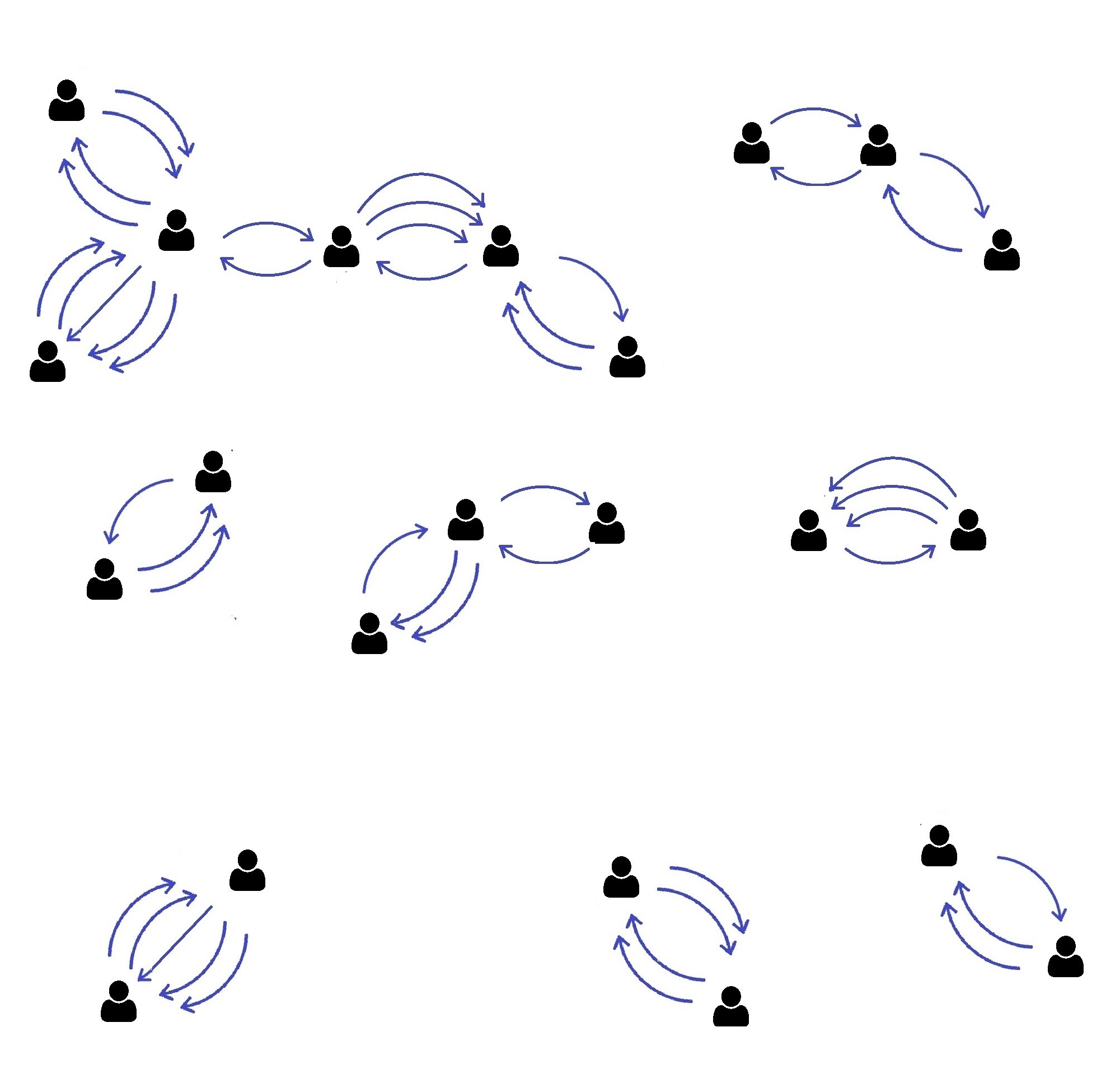}
\label{fig:subfig3}} 
\end{minipage}%
\qquad
\begin{minipage}{0.4\textwidth}%
\centering
\subfloat[Subfigure 4 list of figures text][]{
\includegraphics[width=0.59\textwidth]{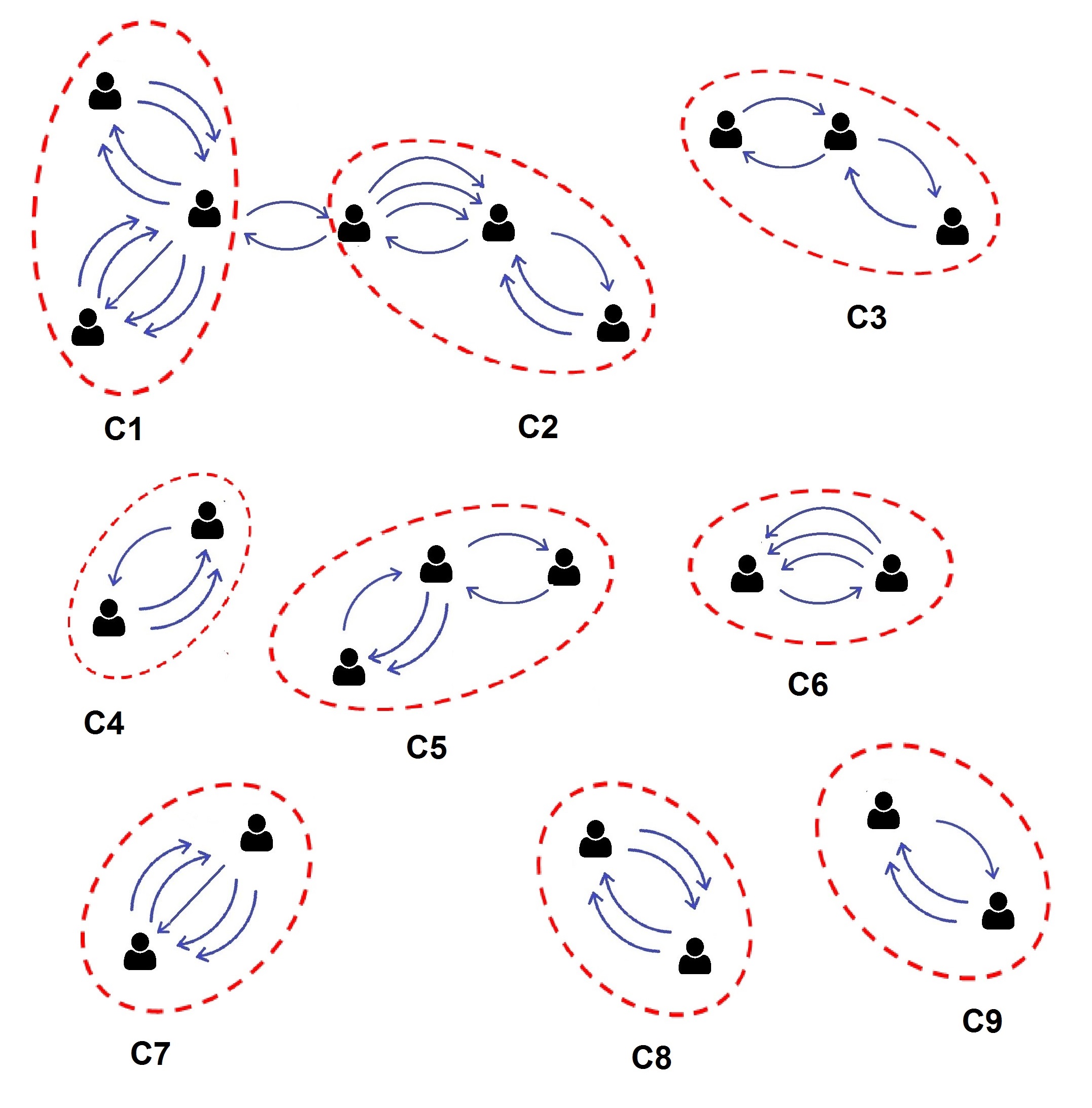}
\label{fig:subfig4}} \\
\subfloat[Subfigure 5 list of figures text][]{
\includegraphics[width=0.59\textwidth]{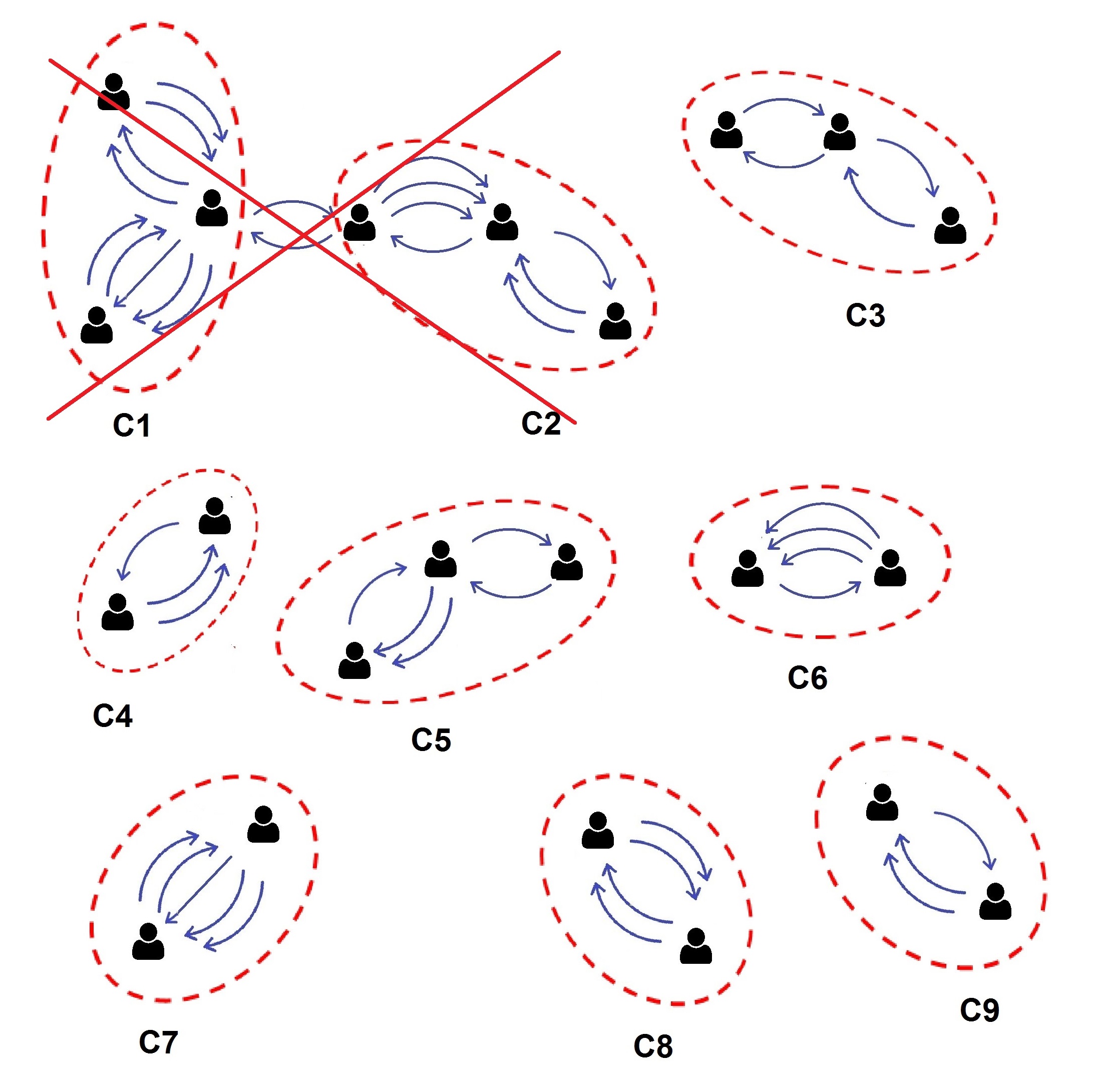}
\label{fig:subfig5}}\\
\subfloat[Subfigure 6 list of figures text][]{
\includegraphics[width=0.59\textwidth]{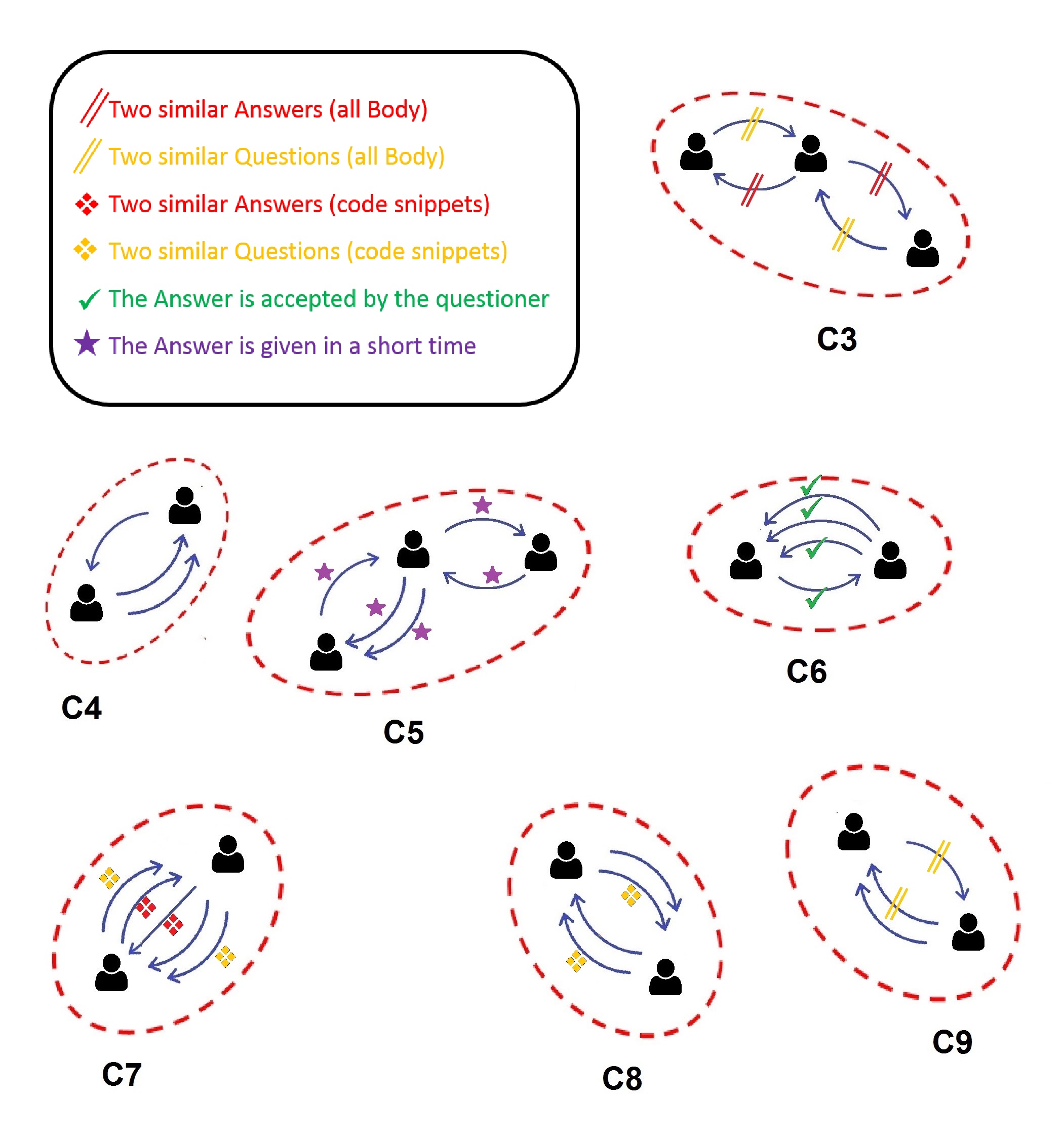}
\label{fig:subfig6}}
\end{minipage}
\qquad
\begin{minipage}{0.4\textwidth}%
\centering
\subfloat[Subfigure 7 list of figures text][]{
\includegraphics[width=0.59\textwidth]{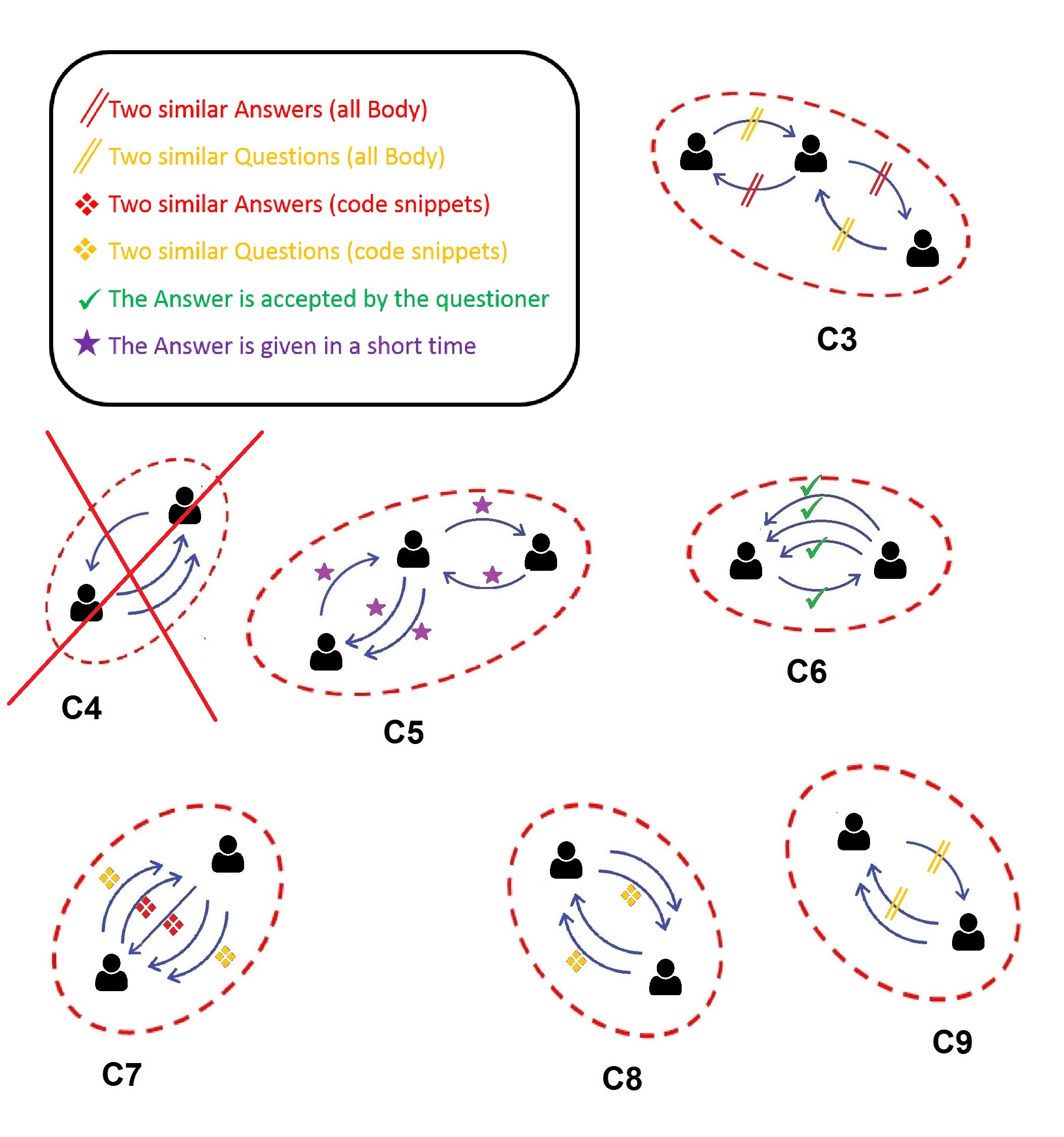}
\label{fig:subfig7}} 
\end{minipage}
\caption{The step-by-step process of selecting suspicious communities.}
\label{fig:algorithm1-step-by-step-description-with-figures}
\end{figure}

\item\bf{Detect Suspicious Community $GC_{V3}$.} Given as input all the communities returned by Louvain algorithm on our Interaction Graph, this step heuristically labels some of the communities as suspicious based on some heuristics which would be explained in the following. We consider two types of similarity, \textit{question similarity} and \textit{answer similarity}. 
We define two thresholds for \bf{question similarity}: 
\begin{inparaenum}[\bfseries (1)]
\item \textit{question body similarity threshold $\tau_{QB}$} based on which two questions are similar if all body (content) of a question is similar to all body (content) of another one, 
\item \textit{question code snippets similarity threshold $\tau_{QC}$} based on which two questions are similar if only the code snippets part of a question is similar to the code snippets part of another one.
\end{inparaenum} 
Also, we define two thresholds for \bf{answer similarity}: 
\begin{inparaenum}[\bfseries (1)]
\item \textit{answer body similarity threshold $\tau_{AB}$} based on which two answers are similar if all body (content) of an answer is similar to all body (content) of another one, 
\item \textit{answer code snippets similarity threshold $\tau_{AC}$} based on which two answers are similar if just code snippets part of an answer is similar to code snippets part of another one.
\end{inparaenum} 
This step tags a community suspicious in two ways. 
First, some of the communities are labeled heuristically as suspicious based on two heuristics:
\begin{inparaenum}[\bfseries (H1)]
\item The community is isolated, i.e., the out degree from a given community is 0. 
\item in the community, there are at least two similar questions asked by the community's users, which have been answered by the other community's users and whose similarity is above $\tau_{QB}$ (or $\tau_{QC}$).
\end{inparaenum} 
Second, some of the communities are labeled heuristically as suspicious based on two heuristics:
\begin{inparaenum}[\bfseries (H1)]
\item The community is isolated, i.e., the out degree from a given community is 0. 
\item in the community, there are at least two similar questions asked by the community's users, which have been answered by the other community's users and whose similarity is above $\tau_{QB}$ (or $\tau_{QC}$) and there are at least two similar answers which have been given by the community's users to questions asked by the other community's users and whose similarity is above $\tau_{AB}$ (or $\tau_{AC}$).
\end{inparaenum}  
To compute similarity, we apply standard preprocessing of Stack Overflow posts, such as removal of HTML tags, separation of textual, code snippets, and hyperlinks. 
We use cosine similarity metric, which is a standard metric to compute textual similarity in software engineering research. Cosine similarity is caluated as:
\begin{equation}
cos(\theta) = \frac{\overrightarrow{a}.\overrightarrow{b}}{||\overrightarrow{a}||||\overrightarrow{b}||} =  \frac{\sum_{1}^n a_ib_i}{\sqrt{\sum_{1}^n a_i^2}\sqrt{\sum_{1}^n b_i^2}}
\end{equation} Here $\overrightarrow{a}.\overrightarrow{b}$ is the dot product of two vectors. In our case, the two vectors are computed as the count frequency of 
individual terms in two documents between which the similarity is computed. Following standard practice, the frequency is normalized using TF-IDF (Term Frequency-Inverse Document Frequency).

\revision{Our algorithm relies on the assumption that two or more people who want to participate in a fraudulent activity respond to each other's questions. Thus, we have selected users who have answered at least one of each other’s questions. We have devised three algorithms based on different assumptions on how the mutual relation forms a suspicious community.}
\revision{We showed the detailed description of algorithm 1 in \fig\ref{fig:algorithm1-detailed-description}.
Also, we presented the step-by-step process of selecting suspicious communities in \fig\ref{fig:algorithm1-step-by-step-description-with-figures}. In \fig\ref{fig:algorithm1-step-by-step-description-with-figures}\subref{fig:subfig1}, we showed a hypothetical interaction graph of users where edges are drawn from question posters to ones who answered those questions. \fig\ref{fig:algorithm1-step-by-step-description-with-figures}\subref{fig:subfig2} is the interaction graph of \fig\ref{fig:algorithm1-step-by-step-description-with-figures}\subref{fig:subfig1}; in this Figure, we could see users and edges with at least one round-robin interaction and deleting the edges and users where all the interactions  are one-directional. The final form of interaction graph of users used in our analysis, where at least one round-robin interaction among users is presented in \fig\ref{fig:algorithm1-step-by-step-description-with-figures}\subref{fig:subfig3}.
\fig\ref{fig:algorithm1-step-by-step-description-with-figures}\subref{fig:subfig4} is the first stage of detecting communities based on round-robin interactions.
In the second stage, we just choose the isolated communities; so you could see the finalized set of isolated communities used in the study in \fig\ref{fig:algorithm1-step-by-step-description-with-figures}\subref{fig:subfig5}. 
The process of analyzing isolated communities based on $GC_{V1}$, $GC_{V2}$, and $GC_{V3}$ has been shown in \fig\ref{fig:algorithm1-step-by-step-description-with-figures}\subref{fig:subfig6}. Finally, you can see the Result of selected isolated communities as suspicious communities by algorithm 1 ($GC_{V1}$, $GC_{V2}$, and $GC_{V3}$) in \fig\ref{fig:algorithm1-step-by-step-description-with-figures}\subref{fig:subfig7}.}
\end{inparaenum}

\subsection{Algorithm 2. Suspicious  User Detection}\label{subsec:algo2-repgamer}
In this section, we describe an algorithm that we developed based on the indicators of reputation frauds that we learned in \sec\ref{subsec:learning-indicators-reputation-fruad}. 
The algorithm detects suspicious users who exhibit unusually high reputation gain in a short time. This algorithm is useful when a reputation fraud may not show any 
distinguishable pattern of being associated to a specific community (i.e., not isolated), rather the fraud may be exploiting the reputation mechanism based on other diverse reputation gaming scenarios including voting rings (e.g., bounty gaming). 

To consider a jump in reputation score as suspicious, we first compute the delta of reputation score 
$\delta_{m,n}^i = \Re_m^i - \Re_n^i$ \revision{of each user $i$ in Stack Overflow between two consecutive dumps denoted by $m$ and $n$.} 
We compute the average of deltas as $\rho_{m,n} = \frac{\sum_{i=1}^N \delta_{m,n}^i}{N}$. To compute the average of deltas, we only consider the users who are active in the last $\tau_{M}$ months. 
We consider the deviation from normality of a user reputation growth as $\Phi_{m,n}^i = \frac{\delta_{m,n}^i - \rho_{m,n}}{\rho_{m,n}}$. 
If $\Phi_{m,n}^i$ is above a given threshold $\tau_{R}$, we label the user as a suspicious.

\subsection{Evaluation of Algorithms}\label{subsec:evaluation-algo-reputation-suspicious}
We evaluate our two proposed algorithms by analyzing the effectiveness of the algorithms to detect suspicious reputations 
in Stack Overflow. We first outline the overall evaluation setup in 
\sec\ref{subsec:evaluation-target}, then present the results of two baselines in \sec\ref{subsec:baseline-fraud-detect}. 
We then present the evaluation results of the two algorithms in \secs\ref{subsec:evaluation-algo1}, \ref{subsec:evaluation-algo2}.

\subsubsection{Experimental Setup}\label{subsec:evaluation-target}
If Stack Overflow team finds a fraudulent activity aimed to increase a user's reputation score, the team removes the score. 
If the user is engaged in various gaming activities, \revision{Stack Overflow may remove the user from Stack Overflow database.} 
Therefore, we cannot identify removed users in Stack Overflow, because they are not available in the Stack Overflow database. 
However, our algorithm may find a suspicious user who is not removed, but one or more of his scores got removed by Stack Overflow. 
Such score removal history is not 
provided in the Stack Overflow data dumps. Therefore, we cannot use the Stack Overflow data dump for our evaluation.

\begin{figure*}[t]
\vspace{-6mm}
	\centering
	\includegraphics[scale=.75]{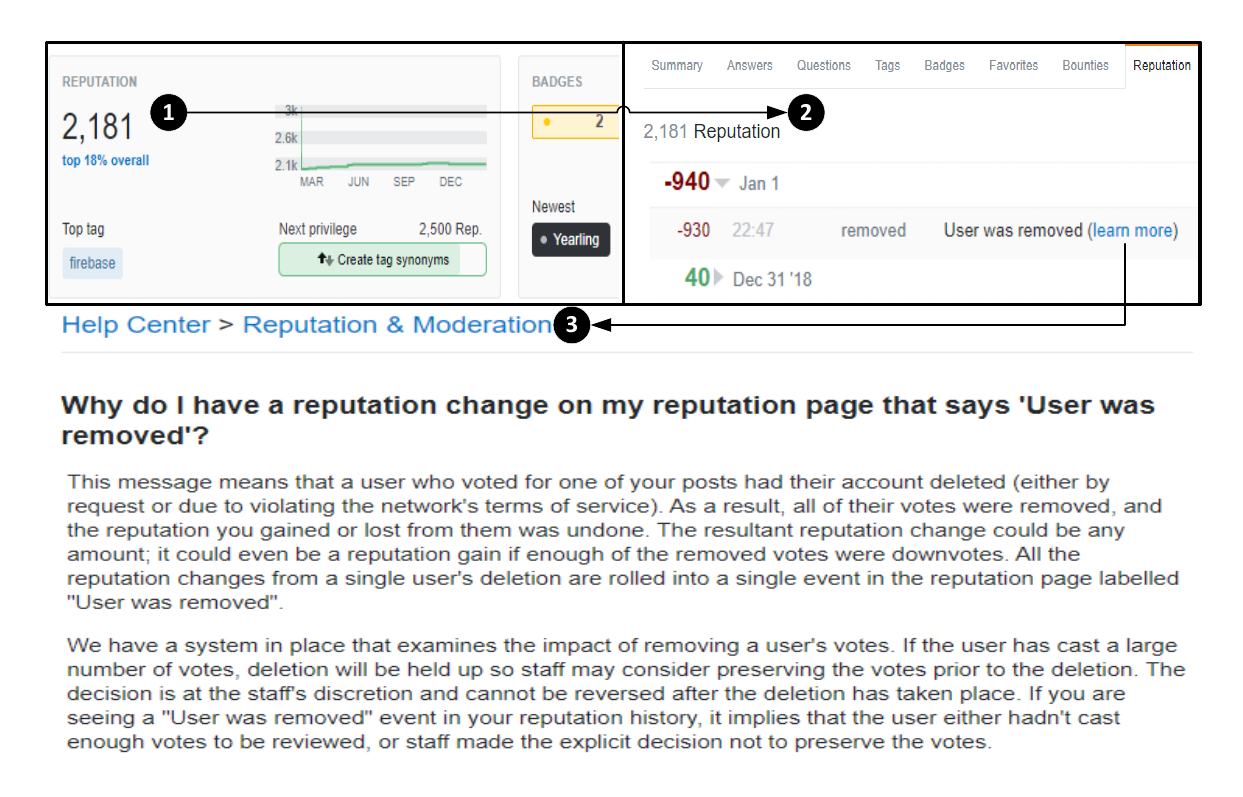}
	\vspace{-6mm}
	\caption{Screenshots of a Stack Overflow fraud user whose scores were removed by Stack Overflow}
	\label{fig:IllustrationScoreRemoved}
\end{figure*}
\begin{figure*}[t]
\vspace{-6mm}
	\centering
	\includegraphics[scale=.95]{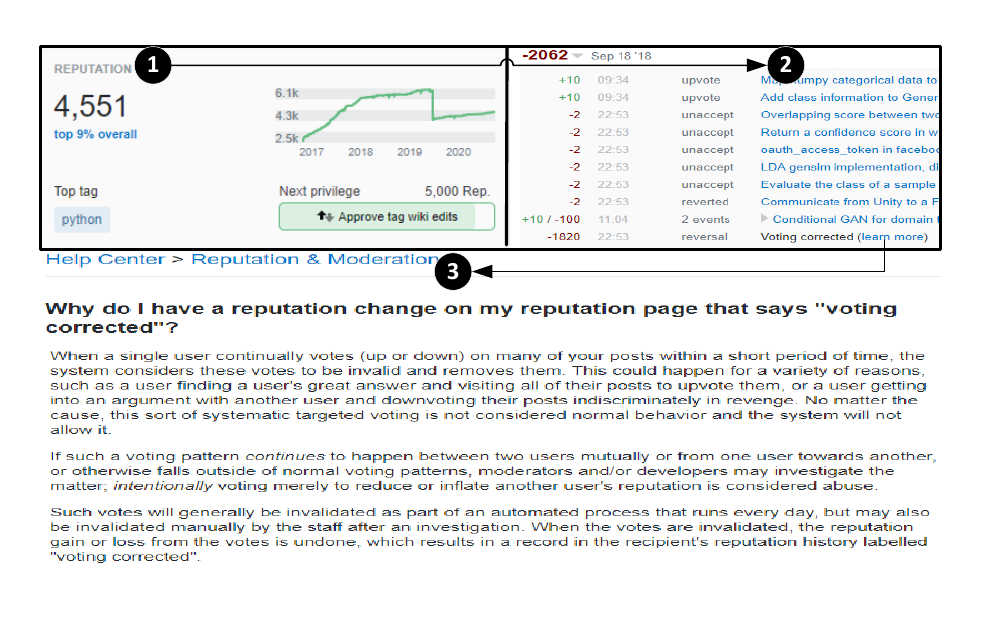}
	\vspace{-10mm}
	\caption{Screenshots of a Stack Overflow fraud user whose scores were reversed by Stack Overflow}
	\label{fig:IllustrationScoreReversed}
\end{figure*}
However, we observed that Stack Overflow online website shows the following types of reputation activities for each user in 
a personalized information dashboard \footnote{\url{https://meta.stackoverflow.com/a/269654}}:
\begingroup
\addtolength\leftmargini{-0.1in}
\begin{quote}
 \revision{\emt{[...] \begin{inparaenum}[(1)]
	\item upvote/downvote (your post was upvoted or downvoted),
	\item unupvote/undownvote (an upvote or downvote previously cast on the post was retracted),
	\item accept/unaccept (your answer was accepted or unaccepted by the person who asked the question, or you accepted an answer on one of your questions),
	\item bounty (you offered a bounty on a question or a bounty was awarded on your answer), 
	\item Removed (indicates the post was deleted; this event is also used when a user is removed (a user who voted for one or more of your posts had their account deleted either by request or due to violating the network's terms of service) \footnote{\label{footnote_UserWasRemoved}\url{https://meta.stackexchange.com/a/126471}}), \item migrated (your post was migrated to another site), \item edit (an edit you suggested to a post or tag wiki was approved), and \item reversal (another user serially voted on your posts, which got reversed)
\end{inparaenum}. [...] }}
\end{quote}
\endgroup

For each activity, the user dashboard shows two fields: \begin{inparaenum}[(1)]
	\item Total scores affected by the activity, and 
	\item Date of activity.
\end{inparaenum} Stack Overflow can decide to `remove' a score of a fraud user based on two types of observations:
\begin{inparaenum}
  \item Event `removed' - User was removed;  \revision{user was removed either through a request or as a consequence of violating the network's terms of service \footref{footnote_UserWasRemoved}}. 
  \item Event `reversal' - Voting corrected; the voting was reversed, but Stack Overflow does not inform whether the voter was removed as well.
\end{inparaenum}
Note that the dashboard does not show who voted for an event. However, the `Removed' event is executed by Stack Overflow only when some votes given towards a user are considered suspicious. As an illustration of a `removed' event, consider the user in \fig\ref{fig:IllustrationScoreRemoved}. The 
box \circled{1} shows that user has a reputation score of 2,181 and the latest badge of the user is ``Yearling'. A `Yearling' is a badge provided to a Stack Overflow user, if he has been active in Stack Overflow for at least one year, earning at least 200 reputation scores. We decided to not show the user name to protect privacy. The box \circled{2} shows the reputation score dashboard of the user. We highlight one 
`Removed' event applied to the user reputation on January 1, 2018. A total of 940 scores of the user were removed. The reason given by Stack Overflow was `User was removed'. The bottom box in \fig\ref{fig:IllustrationScoreRemoved} which is circled as \circled{3} shows a screenshot 
of Stack Overflow help page. The page explains the reason behind the removed event. It shows that the 940 scores were removed, because the users who voted for the score were deleted from Stack Overflow. The second paragraph of \circled{3} offers further explanation that such deletion typically happens, when a user has cast large number of votes towards another user. This means, the voting was considered as suspicious. 
The paragraph also explains that such voting may happen in multiple posts, i.e., the two users could be engaged in fraud scenarios in multiple posts.
As an illustration of a `reversal' (i.e., voting corrected) event, consider the user in \fig\ref{fig:IllustrationScoreReversed}. The box \circled{1} shows a user with reputation 
score of 4,551 and the box \circled{2} shows the reputation history of the user on September 18, 2018. It shows that the user got an increase of 30 in his reputation score (through 3 upvotes), but 
the user also got a total of 1,932 scores dropped on that day. Out of those 1,932 dropped scores, the 1,820 scores were dropped due to a `reversal' event. The explanation from Stack Overflow 
is that the voting was corrected. The box \circled{3} explains that a voting is corrected when a single user continuously votes many posts of another same user in a very short time. 

Therefore, if our algorithm identifies a user as suspicious, we 
can verify by looking at his reputation score dashboard, given that Stack Overflow was able to identify the fraudulent activity. 
If his score was removed/reversed by Stack Overflow, we can consider that our algorithm was successful in detecting the reputation fraud. 

As we noted above, to learn about the reputation history of a user for our evaluation, we have to scrape the online Stack Overflow website. 
The reputation history of each user can have hundreds of web pages, each generated dynamically (using AJAX). Scraping dynamic web pages requires the use of online web application automated testing framework (e.g., Selenium web driver), which can load a dynamically-generated web page (if proper address is given). 
Given that our algorithms run against all users in Stack Overflow data dumps, we thus needed to develop a script to download the reputation history of a user from online Stack Overflow website. While Stack Overflow did not block us from downloading the reputation history using our Selenium-based dynamic web page scraper, it imposed strict rate limits on downloads. Stack Overflow does this to ensure that automated bots do not impact the website performance. 
This limit significantly slowed down the download process, 
often taking days for just a single user.  
Therefore, we selectively downloaded subset of Stack Overflow users 
for our evaluation as follows. 

When the algorithm returns less than 200-300 users, we attempt to download the reputation history of all such others. Otherwise, 
we download the reputation history of a statistically significant subset of the returned users. For example, 
with a 99\% confidence level and 10 interval, this will be 166 from around 200,000 users.  
Given that each such user is considered as a potential suspicious reputation by our algorithm, the analysis of such users can only provide information of the precision 
(i.e., positive predictive value) of the algorithm. To compute recall (i.e., true positive rate), we ideally require to analyze the reputation history of all the millions of users. 
Given that this is not possible, we compute `relative recall' by following Sampson et al.~\cite{Sampson-RelativeRecall-BMC2006}, i.e., by 
analyzing statistically significant samples of the users who are not considered as suspicious by our algorithm. 
\revision{For this purpose, we randomly selected a set of users and looked at their reputation scores' history thoroughly. For knowing how many users we should pick randomly, We used a 95\% confidence level with 10 intervals, which would give us 84 Stack Overflow users. In 84 users that we thoroughly examined, 10 of them had `reversal' and `removed' tags.  These tags are usually used when suspicious activities lead to a decrease in reputation scores. Thus, we labeled them as fraudulent users.}
Based on this setup, we report four 
evaluation metrics (Precision $P$, Relative Recall $R$, F1-score $F1$, and Accuracy $A$) as follows:  

{\scriptsize
\begin{eqnarray*}
P  = \frac{TP}{TP+FP},~
R = \frac{TP}{TP+FN},~
F1 = 2*\frac{P*R}{P+R},
A = \frac{TP+TN}{TP+FP+TN+FN}
\end{eqnarray*}}
TP = Correctly classified as a fraud, 
FP = Incorrectly classified as a fraud, TN = Correctly classified as not a fraud, 
FN = Incorrectly classified as not a fraud.

%


\nd\bf{Evaluation Corpus.} We used the Stack Overflow database dump provided by SOTorrent~\cite{Baltes-SOTorrent-MSR2018}, which 
is created from Stack Overflow original data dump by analyzing post versions. The SOTorrent 
datasets are also shared in Google BigQuery dataset. This is useful for our 
algorithm 2, which checks the change in reputation scores for a user between two dates. 
For Algorithm 1, we used the SOTorrent dataset of September 2018~\cite{website:sotorrent-23092018}. This was the most recent data dump available during 
the start of our analysis of this paper. For Algorithm 2, to 
analyze jumps in reputation scores, we analyze three SOTorrent datasets: \begin{inparaenum}
\item September 2018~\cite{website:sotorrent-23092018},
\item December 2018~\cite{website:sotorrent-09122018}, and
\item March 2019~\cite{website:sotorrent-17032019}.
\revision{The two algorithms 1 and 2 are completely different and have no relationship with each other.  Algorithm 1 investigates the connections among users who may form a voting ring; thus, uses a single dump. In contrast, Algorithm 2 explores the difference between reputation jumps and downfalls. Therefore,  we have to assess two consecutive dumps to find suspicious users. Stated differently, Algorithm 1 explores the connections among users whereas Algorithm 2 assesses the difference between jumps and downfalls of the reputation of a single user. Also, we are not tracking the same user with the two algorithms 1 and 2. As a result, the fact that the two dumps are likely to contain different users should not affect our results.
Based on the assumption of Algorithm 2, we only had to look deeply into consecutive dumps of SO.  Three-month time interval is selected due to the time difference between how long it takes for SO to release consecutive dumps and no further reason was involved.  Selecting larger time intervals must be based on multiples of three-month intervals that Stack Overflow releases dumps. Further discussion on evaluating different time windows is presented in \sec\ref{sec:threats}. }

\end{inparaenum}

\subsubsection{Baseline}\label{subsec:baseline-fraud-detect}
We compare our algorithms against two baselines, both based on `change in reputation scores'. 
The first baseline analyzes users whose reputation scores showed an upward jump in two Stack Overflow data dumps: September 2018 and December 2018. 
We call this baseline $B_{U}$. The second baseline analyzes users whose reputation scores dropped in the most recent data dump compared to the previous data dump. We call this 
baseline $B_{D}$. For both baselines, we only consider Stack Overflow users whose change in scores is more than the average. This means for $B_U$, 
we analyze users whose jump in scores is more than the average jump in user scores between the two dumps. Note that this rationale is based on the study of reputation fraud indicators we conducted in \sec\ref{subsec:learning-indicators-reputation-fruad}. Given that a reputation fraud user is focused on increasing his score in 
a short time, the change in his score between two dumps could be more than the normal (i.e., average) jump in reputation score in the two dumps. For $B_D$, we only analyze 
the users whose drop in scores is more than the average drop in users (out of all user whose scores were dropped) between the two dumps. This baseline is the opposite of 
$B_U$. Here, we are interested in the fraud users for whom we could not see a jump in reputation score, because Stack Overflow has reduced their scores for suspicious activities. 
\revision{In other words, to detect new suspicious users, we hypothesized that the users that StackOverflow has detected as fraudulent have recognizable drops in their reputation.}
We only look at the users whose scores dropped below average to ensure that we avoid flagging a user as suspicious while his score was dropped due to normal reasons. For example, the rollback 
of a previously approved edit will deduct a -2 score from a user. However, that action is highly unlikely to be considered as fraud. 

To calculate the average in jump or drop, we only considered users who were active between the two dumps, i.e., if a user was not active between the two dumps we did not include 
his score in the average calculation. Thus the two baselines look at users whose growth (jump or drop) in reputation scores 
is more than the average reputation growth in active users that participated in Stack Overflow discussions by asking or answering to questions.   

In our two above data dumps, a total of 660,092 users showed an increase and 41,005 users showed a decrease in reputation score. 
We evaluate each baseline as follows: 
\begin{inparaenum}
\item We create a list of all users returned by the baseline, i.e., for $B_U$, all users whose jump in scores is above the mean as defined above and for $B_D$ all users whose drop in scores 
is below the mean as defined above. 
\item From the list in the first step, we randomly pick 100 users. This sample is statistically significant with a 99\% confidence level and 10 interval.
\item For each of the 100 users, we investigate the reputation history of the user between the time of the two data dumps (i.e., September - December 2018). If we see one or more 
`removed'/`reversal' event of the user at the time of analysis, we consider that our baseline is successful to identify the fraud. 
\end{inparaenum} In \tbl\ref{tab:perf-baseline}, we show the performance of the two baselines. The $B_{D}$ shows better performance than $B_{U}$, 
i.e., users whose reputation scores got dropped beyond the average drop in scores are more likely to be suspicious than the users whose 
reputation scores got dropped beyond the average jump in scores. For $B_D$, the majority of misclassifications with respect to Stack Overflow actions happened due to the following reasons:
\begin{inparaenum}
\item Bounty. The user has awarded bounties to other users. In Stack Overflow, users often offer bounties to encourage quick but quality answers.
\item Downvote. The user was subject to downvotes. That means the user could have posted some below quality questions/answers. This also means that user could have been subject to 
revenge voting from other users (i.e., reputation frauds).  
\end{inparaenum} For the $B_U$, most of the misclassified frauds appeared to gain their reputations by being more active than normal users and their activity did not raise any concern of fraudulent activities.
\revision{Reputation gain is so prevalent that finding a suspicious user based on reputation gain is not practical. Thus, we look at all the users altogether. The majority of them gained their reputations through legitimate reputation improvement scenarios that SO encourages.
This means only a relatively small growth in reputation may occur by fraudulent activities.  As a result, 0.07 matched our expectation.}

\begin{table}[t]
  \centering
  \caption{The performance of baselines to detect suspicious reputations}
    \begin{tabular}{lrrrr}\toprule
    \textbf{Baseline} & {\textbf{Precision}} & {\textbf{Recall}} & {\textbf{F1-Score}} & {\textbf{Accuracy}} \\ \midrule
    $B_{D}$ (Dropped score) & 0.30  & 0.75  & 0.43  & 0.57 \\
    $B_{U}$ (Jumped score)  & 0.07  & 0.41  & 0.12  & 0.44 \\ \bottomrule
    \end{tabular}%
  \label{tab:perf-baseline}%
\end{table}%

\subsubsection{Performance of Algorithm 1 - Suspicious Community}\label{subsec:evaluation-algo1}
We ran our algorithm to detect suspicious communities in 
SOTorrent dataset of September 2018~\cite{website:sotorrent-09122018}. 
Recall that for the three versions of our proposed algorithm 1, we use four types of metrics while making a decision. 
\begin{enumerate}
	\item $N_{L}$: Number of links between users in a community ($GC_{V1, V2}$) 
	\item $T$: Time taken to answer to a question (for $GC_{V1}$)
	\item $S_{Q}$: Similarity in posted questions in a community ($GC_{V3}$). 
	\item $S_{A}$: Similarity in posted answers in a community ($GC_{V3}$). 
	\end{enumerate}
	For both $S_{Q}$ and $S_{A}$, two types of similarity were considered: \begin{inparaenum}[(a)]
	\item $B$: all post contents, and 
	\item $C$: only code snippets in the post. 
	\end{inparaenum}

\begin{table}[t]
  \centering
  \caption{The different configurations (cases) of Algorithm 1}
    \begin{tabular}{rllrr}\toprule
    \multicolumn{1}{l}{\textbf{Algorithm}} & \textbf{Case} & \textbf{Metrics} & \multicolumn{1}{l}{\textbf{\#Community}} & \multicolumn{1}{l}{\textbf{\#User}} \\ 
    \midrule
    \multicolumn{1}{l}{$GC_{V2}$} & C1    & $N_{L} \geq (\tau_{L}=2)$ & 2024  & 4115 \\
          & C2    & $N_{L} \geq (\tau_{L}=6)$ & 157   & 352 \\
          & C3    & $N_{L} \geq (\tau_{L}=8)$ & 66    & 148 \\
          & \bf{C4}    & $N_{L} \geq (\tau_{L}=10)$ & 29    & 69 \\
          \midrule
    \multicolumn{1}{l}{$GC_{V1}$} & {C5}    & $N_{L} \geq (\tau_{L}=6),  T \leq (\tau_{T}=24H)$ & 83  & 234 \\
          & C6    &  $N_{L} \geq (\tau_{L}=6),  T \leq (\tau_{T}=1H)$ & 10  & 27 \\
          & C7    &  $N_{L} \geq (\tau_{L}=6),  T \leq (\tau_{T}=0.5H)$ & 7  & 20 \\
          & C8    &  $N_{L} \geq (\tau_{L}=6),  T \leq (\tau_{T}=0.25H)$ & 4  & 10 \\
          & \bf{C9}    &  $N_{L} \geq (\tau_{L}=8),  T \leq (\tau_{T}=24H)$ & 37  & 111 \\
          & C10    &  $N_{L} \geq (\tau_{L}=8),  T \leq (\tau_{T}=1H)$ & 3  & 8 \\
          & C11    &  $N_{L} \geq (\tau_{L}=8),  T \leq (\tau_{T}=0.5H)$ & 3  & 8 \\
          & C12    &  $N_{L} \geq (\tau_{L}=8),  T \leq (\tau_{T}=0.25H)$ & 1  & 2 \\
          \midrule
    \multicolumn{1}{l}{$GC_{V3}$} & {C13}    & $S_{QB} \geq (\tau_{QB}=0.89)$ & 97    & 229 \\
          & \bf{C14}   & $S_{QB} \geq (\tau_{QB}=0.95)$ & 33    & 81 \\
     & C15 & $S_{QC} \geq (\tau_{QC}=0.80)$ & 92    & 303 \\
    & \bf{C16} & $S_{QC} \geq (\tau_{QC}=0.90)$ & 39    & 99 \\
          & C17   & $S_{QB} \geq (\tau_{QB}=0.89),  S_{AB} \geq (\tau_{AB}=0.89)$ & 14    & 38 \\
          & \bf{C18}   & $S_{QB} \geq (\tau_{QB}=0.95),  S_{AB} \geq (\tau_{AB}=0.95)$ & 6     & 16 \\
          & {C19}   & $S_{QC} \geq (\tau_{QC}=0.80),  S_{AC} \geq (\tau_{AC}=0.80)$ & 17     & 64 \\
          & \bf{C20}   & $S_{QC} \geq (\tau_{QC}=0.90),  S_{AC} \geq (\tau_{AC}=0.90)$ & 7     & 28 \\
          \bottomrule
    \end{tabular}%
  \label{tab:algo1-cases-stat}%
\end{table}%
	
In \tbl\ref{tab:algo1-cases-stat}, we show different cases of each version of our Algorithm 1, where each case corresponds to specific 
values of the metrics above. The first column (`Algorithm') denotes the versions of the algorithm. The second column (`Case') denotes 
the different cases for each version that we ran on our database. The third column (`Metrics') shows the specific metric value used for a case. 
The last two columns (`\#Community' and `\#User') present statistics on the number of communities and users the algorithm returned as potentially suspicious (i.e., fraud). 
As the value of a metric increases, the number of communities (and users) decreases.  
 
We analyze six highlighted cases in \tbl\ref{tab:algo1-cases-stat} to compute the performance of Algorithm 1 (bold in Case 
column of \tbl\ref{tab:algo1-cases-stat}). 
The cases are selected for the following reasons: 
\begin{inparaenum}
  \item The more links existed between users in an isolated community in which all users have accepted each others' answers, the more engaged those users are to increase their reputation score within the community (C4).
  \item A suspicious community could be highly engaged but isolated within  short time to answer each others questions  (C9).
  \item Users in a suspicious community can post similar questions and answers multiple times (C14, C16, C18, C20).
\end{inparaenum}
\begin{table}[t]
  \centering
  \caption{Performance of Algorithm 1 Cases as defined in \tbl\ref{tab:algo1-cases-stat}}
    \begin{tabular}{lrrrrr}\toprule
    \textbf{Case} & {\textbf{\#User}} & {\textbf{Precision}} & {\textbf{Recall}} & {\textbf{F1-Score}} & {\textbf{Accuracy}} \\ \midrule
    C4    & 64    & 0.69  & 0.81  & 0.75  & 0.80 \\
    C9    & 111   & 0.73  & 0.89  & 0.80  & 0.79 \\
    C14   & 81    & 0.41  & 0.77  & 0.53  & 0.65 \\
    C16   & 99    & 0.46  & 0.82  & 0.59  & 0.65 \\
    C18   & 16    & 0.62  & 0.50  & 0.56  & 0.84 \\
    C20   & 28    & 0.54  & 0.60  & 0.57  & 0.79 \\
    \midrule
    Total/Avg & 399   & 0.58  & 0.73  & 0.63  & 0.75 \\ 
    \bottomrule
    \end{tabular}%
  \label{tab:algo1-perf}%
\end{table}%
In \tbl\ref{tab:algo1-perf}, we show the overall performance of our algorithm for each selected case from \tbl\ref{tab:algo1-cases-stat}.
Each case performed better (F1-score) than the two baselines. 
Among the six cases, the case C9 showed the best performance. This case looks at isolated communities of users with at least 8 edges, which shows that the community's users have a lot of communications with each other and answer each other's questions in a very short time (in less than 24 hours).
With regards to precision, among these two cases (C4, C9), a combination of having a lot of communications and answering in a short time (i.e., C9) is more effective than a combination of having a lot of communications and accepting all answers (i.e., C4).
\revision{Proper use of parameters that restrict the search lead to detect more probable suspicious users and fewer false alarms. As an illustration, we restricted the time to find users in 24 Hours rather than 36 Hours or 48 Hours. If we did not restrict the parameter, we would find more users but less likely to be suspicious. Also, the analysis would be more time-consuming. Furthermore, our purpose was to show the possibility of the existing fraud in the platform not to detect all types of fraud in the platform. By these restrictions, we found suspicious cases which Stack Overflow team did not detect and the team acknowledged they were suspicious.}
The four cases (C14, 16, 18, and 20) analyze the similarity of the posted contents in questions and answers. With regards to precision, among these four cases, the similarity in posted questions and posted answers (C18, and 20) seems like a better indicator than the similarity in posted questions (C14, 16).

\begin{table}[t]
  \centering
  \caption{Basic statistics of detected suspicious communities based on various runs of different versions of suspicious community detection based on Algorithm 1}
  \begin{tabular}{lllllllll}
    \toprule
    \rotatebox[origin=c]{60}{\bf{Algorithm}} &
    \rotatebox[origin=c]{60}{\bf{Case}} &
    \rotatebox[origin=c]{60}{\bf{\#Com}} &
    \rotatebox[origin=c]{60}{\bf{Ave(\#Node)$\pm$ S.D.}} &
    \rotatebox[origin=c]{60}{\bf{Min(\#Node)}} &
    \rotatebox[origin=c]{60}{\bf{Max(\#Node)}} &
    \rotatebox[origin=c]{60}{\bf{Ave(\#Edge)$\pm$ S.D.}} &
    \rotatebox[origin=c]{60}{\bf{Min(\#Edge)}} &
    \rotatebox[origin=c]{60}{\bf{Max(\#Edge)}} \\
    \midrule
    $GC_{V2}$ & C4  & 29 & $2.38 \pm 0.56$ & 2 & 4 & $14.31 \pm 6.46$ & 10 & 33 \\
    $GC_{V1}$ & C9  & 37 & $3 \pm 1.13$ & 2 & 6 & $10.54 \pm 3.18$ & 8 & 21 \\
    $GC_{V3}$ & C14 & 33 & $2.45 \pm 1.62$ & 2 & 11 & $7.06 \pm 12.03$ & 2 & 71 \\
    $GC_{V3}$ & C16 & 39 & $2.54 \pm 1.52$ & 2 & 11 & $8.44 \pm 11.59$ & 2 & 71 \\
    $GC_{V3}$ & C18 & 6  & $2.67 \pm 1.03$ & 2 & 4 & $8.33 \pm 5.28$ & 2 & 15 \\
    $GC_{V3}$ & C20 & 7  & $4 \pm 3.21$ & 2 & 11 & $18.71 \pm 23.66$ & 2 & 71 \\
    \bottomrule
  \end{tabular}
  \label{tab:detailed-statistics-of-the-communities}
\end{table}

\revision{In \tbl\ref{tab:detailed-statistics-of-the-communities}, we show some statistics about the six highlighted cases in \tbl\ref{tab:algo1-cases-stat}. 
Based on different configurations, \tbl\ref{tab:detailed-statistics-of-the-communities} shows the number of communities detected by our algorithms, as well as the minimum, maximum, and average \(\pm\) standard deviation numbers of nodes and edges for all those communities.
For instance, 29 communities that are detected by algorithm $GC_{V2}$ with C4 have $2.38$ nodes and $14.31$ edges on average and have $0.56$ Standard Deviation of nodes and $6.46$ Standard Deviation of edges. One may ask why the number of detected communities is not large and why the communities do not have many users. Since we did not have insider information, we took a very conservative approach and were restrictive. C4 considers communities having at least $10$ links and all answers in the communities have been accepted by the members of those communities. C4 has only $29$ communities, all of which have $69$ users in total.} 
If you look at \tbl\ref{tab:algo1-cases-stat}, C1 shows communities having at least $2$ links and all answers in this community have been accepted by the members of that community. Case C1 could detect around $2000$ communities, all of which have around $4100$ users in total.
\revision{For detailed statistics of the communities, please go to the 
\emph{"additional\_info\_about\_Table 2.rar"} file that is contained in the folder \emph{"Additional Info for TOSEM"} located at: \url{https://github.com/mazloomzadeh/Reputation-Gaming.git}.
You can find all the information about the number of nodes and number of edges of each community listed in \tbl\ref{tab:detailed-statistics-of-the-communities} within the folder named \emph{"suspiciousCommunities\_numberOfnode\_numberOfedge"}.
Also, you could find some statistics about the nodes and edges of each community in the file \emph{"statistics\_of\_SuspiciousCommunities\_algorithm1.xlsx"}.}

\revision{In \tbl\ref{tab:detailed-statistics-of-userRep-Algo1}, we presented more statistics about users' reputations in the six cases of algorithm 1 which is highlighted in \tbl\ref{tab:algo1-cases-stat}.}

\begin{table}[t]
  \centering
  \caption{Basic statistics of detected suspicious users' reputations based on various runs of different versions of suspicious community detection based on Algorithm 1}
  \begin{tabular}{llllllll}
    \toprule
    \rotatebox[origin=c]{0}{\bf{Algorithm}} &
    \rotatebox[origin=c]{0}{\bf{Case}} &
    \rotatebox[origin=c]{0}{\bf{Ave(Rep) $\pm$ S.D.}} &
    \rotatebox[origin=c]{0}{\bf{Median(Rep)}} &
    \rotatebox[origin=c]{0}{\bf{Variance(Rep)}} &
    \rotatebox[origin=c]{0}{\bf{Min(Rep)}} &
    \rotatebox[origin=c]{0}{\bf{Max(Rep)}} \\
    \midrule
    $GC_{V2}$ & C4  & $1405.41 \pm 2789.53$ & 664 & 7781453.51 & 35 & 21338 \\
    $GC_{V1}$ & C9  & $5623.99 \pm 13833.9$ & 1590 & 191376913.1 & 45 & 120604 \\
    $GC_{V3}$ & C14 & $1142.74 \pm 3216.9$ & 299 & 10348441.79 & 15 & 22983 \\
    $GC_{V3}$ & C16 & $1322.51 \pm 3048.31$ & 395 & 9292175.6 & 15 & 22983 \\
    $GC_{V3}$ & C18 & $992.69 \pm 1169.11$ & 443 & 1366827.56 & 29 & 3502 \\
    $GC_{V3}$ & C20 & $1346.18 \pm 2750.98$ & 383 & 7567915.78 & 15 & 14064 \\
    \bottomrule
  \end{tabular}
  \label{tab:detailed-statistics-of-userRep-Algo1}
\end{table}

\begin{table}[t]
  \centering
  \caption{Coverage of reputation frauds (G) found by Stack Overflow in our detected suspicious communities (\#Com)}
    \begin{tabular}{lllllll}\toprule
    \bf{Case} & \bf{\#Com} & \bf{$\geq$1G} & \bf{$100\%$G}   & \bf{$\geq 75\%$G} & \bf{$\geq 50\%$G} & \bf{$<50\%$G} \\ 
    \midrule
    C4   & 29 & 86.2\% & 34.5\% & 37.9\% & 86.2\% & 13.8\% \\
    C9   & 37 & 81.1\% & 54.1\% & 62.2\% & 78.4\% & 21.6\% \\
    C14  & 33 & 63.6\% & 15.2\% & 18.2\% & 60.6\% & 39.4\% \\
    C16  & 39 & 69.2\% & 20.5\% & 25.6\% & 61.5\% & 38.5\% \\
    C18  & 6 & 66.7\% & 33.3\% & 50.0\% & 66.7\% & 33.3\% \\
    C20 & 7  & 85.7\% & 42.9\% & 57.1\% & 71.4\% & 28.6\% \\
    \bottomrule
    \end{tabular}%
  \label{tab:coverage-gamer-in-community}%
\end{table}%

In \tbl\ref{tab:coverage-gamer-in-community}, we show the coverage of suspicious reputation frauds that we could assess using Stack Overflow in each community detected by the six cases. 
We break down the coverage in multiple columns: \begin{inparaenum}[(1)]
\item $\ge1G$: shows the percentage of the suspicious community with at least one probable reputation fraud.
\item $100\%G$: shows the percentage of suspicious community with all the community members likely had reputation gaming.
\item $\ge75\%G$, $\ge50\%G$, $<50\%G$: show the percentage of suspicious community with at least 75\%, 50\% and less than 50\% community members likely had reputation frauds, respectively.
\end{inparaenum} 
\revision{For instance, case C4 in \tbl\ref{tab:coverage-gamer-in-community} could find $29$ suspicious communities. We break down the coverage of this case in multiple columns:
\begin{inparaenum}[(1)]
\item $\ge1G$: shows that at least one member of $34.5\%$ of suspicious communities detected by C4 had reputation fraud \item $100\%G$: shows that all the members of $34.5\%$ of suspicious communities detected by C4 had reputation fraud. \item $\ge75\%G$: shows that at least $75\%$ of members of $37.9\%$ of suspicious communities detected by C4 had reputation fraud, $\ge50\%G$: shows that at least $50\%$ of members of $86.2\%$ of suspicious communities detected by C4 had reputation fraud, $<50\%G$: shows that less than $50\%$ of members of $13.8\%$ of suspicious communities detected by C4 had reputation fraud.
\end{inparaenum}}
Among the six cases, C9 (minimum link $\ge 8$ and time of response $\le 24H$) has the maximum community with 100\% coverage. However, 
the cases C4 (minimum link $\ge 10$ and all answers have been accepted) and C20 (similarity in questions' content and in answers' content) show the most coverage with communities containing at least one suspicious reputation. Given that 
at least one member in those communities is found violating Stack Overflow network rules and given that the members in the community show high degree of communication, it 
could either denote that Stack Overflow detection script may have missed those other users or the Stack Overflow algorithm relies on more information than us to 
make a final decision on those users.  
\begin{table}[t]
  \centering
  \caption{Percentage of suspicious communities (as assessed by Stack Overflow) with at least one probable fraud user whose reputation scores were removed during and after the formation of the community}
    \begin{tabular}{lrlllll}\toprule
    \textbf{Case} & \multicolumn{1}{l}{\textbf{\#Com}} & \textbf{During} & \textbf{In 1D} & \textbf{In 7D} & \textbf{In 14D} & \textbf{In 30D} \\
    \midrule
    C4    & 25    & 76\%  & 76\%  & 76\%  & 80\%  & 80\% \\
    C9    & 30    & 80\%  & 80\%  & 83\%  & 87\%  & 87\% \\
    C14   & 21    & 52\%  & 57\%  & 57\%  & 62\%  & 62\% \\
    C16   & 27    & 59\%  & 67\%  & 67\%  & 70\%  & 70\% \\
    C18   & 4     & 75\%  & 75\%  & 75\%  & 75\%  & 75\% \\
    C20   & 6     & 83\%  & 100\% & 100\% & 100\% & 100\% \\
    \bottomrule
    \end{tabular}%
  \label{tab:comm-proximity}%
\end{table}%
In \tbl\ref{tab:comm-proximity}, we compute the \it{proximity} of time ($T_1$) of formation of a community where we could verify at least one suspicious fraud using 
Stack Overflow reputation dashboard and the time ($T_2$) when the score of at least one likely reputation fraud in the community was removed by Stack Overflow. Time $T_1$ 
uses two time frames, the time $T_{1,Q}$ when the first question in the community was posted and the time $T_{1,A}$ when the last answer in the community was posted. 
If a score of a user in a community got removed before the formation of the community, we do not include that removal event into this analysis.  
The column `During' in \tbl\ref{tab:comm-proximity} shows the percentage of the likely gaming communities that had the score of at least one user got removed by Stack Overflow 
during the formation of the community. The columns `In 1D, 7D, 14D, 30D' denote the percentage of the suspicious communities where the score of one or more 
user got removed by Stack Overflow within 1, 7, 14, and 30 days of the formation of the community. For example, `In 1D' means within 24 hours after $T_{1,A}$. 
The three cases (C4, C9, C20) show between 76\%-83\% coverage, i.e., Stack Overflow actively identified those suspicious users during 
the formation of the communities. The saturation points in the coverage for all cases reach in two weeks (i.e., 14D) after the formation of the communities, i.e., after that we do not see Stack Overflow reducing the score of any more user in those communities.

\subsubsection{Performance of Algorithm 2 - Suspicious User}\label{subsec:evaluation-algo2}
Our proposed algorithm 2 detects suspicious users as the Stack Overflow users that showed an \it{unusual} increase in their reputation scores in a short time. 
Departure from normality for a user $i$ is computed using the metric $\phi_{m,n}^i$ between two consecutive Stack Overflow datasets $m$ and $n$, which tells 
the extent of reputation growth of the user beyond the average reputation growth of all \it{active} users between the two datasets. An \it{active} user is defined 
as a user who accessed Stack Overflow as a registered user during the creation time of the two datasets. 
In \tbl\ref{tab:algo2-growth-threshold} we show three threshold values for $\phi_{m,n}^i$, for which we ran our algorithm. The first threshold (i.e., C1) 
points us to the users whose reputation growth is more than the average growth by twice the standard deviation of the growth, i.e., a user is considered a suspicious user if 
his reputation growth between two datasets is at least $\mu + 2*stddev$. In the three datasets, the average growth (i.e., $\mu$) is around $15$ and the standard deviation (i.e., stddev) 
is around $220$. The high standard deviation is due to users above the 90th percentile with scores much more than the rest of the users. Indeed, the reputation growth of the users around the 99th percentile (Case C2) is around 65 times more than the average growth and the users around the 100th percentile (Case C3) show an incredible at least $130$ times more growth than the average growth. \revision{A subtle measure of outlier for data with normal distribution is $\mu + 3*stddev$. That is, if we know the distribution is indeed Normal Gaussian. First of all, the distribution of the dataset has not been investigated.  Moreover, in the case of Gaussian distributions, the $\mu + 2*stddev$ covers $95\%$ of data so data outside the measure may be considered suspicious by some means.  Again, we are not aware of what the distribution of the data is.  Thus, the use of $\mu + 2*stddev$ may not lead to cover $95\%$ of data as is the case with normal distribution, or $\mu + 3*stddev$ also may not actually cover $99.7\%$ of data as it covers in a normal distribution.
The users that we have detected in cases C2 and C3 in \tbl\ref{tab:algo2-growth-threshold} are more restrictive than what a person has suggested.  Since $\mu + 3*stddev$, where the $\mu$ is around $15$ and the standard deviation is around $220$ the upper limit adds up to $675$. In cases of C2, we look for $65$ times the mean, which is almost equal to $65*15 = 975$, and for C3 we search $130$ times the mean which is equal to $1950$. Except for case C1 which points us to the users whose reputation growth is more than the average growth by twice the standard deviation of the growth (that $\mu + 2*stddev$), in both C2 and C3 scenarios, the search is much more restrictive than $675$. Other cases (C2 and C3) point to users that have been presented in \tbl\ref{tab:algo2-growth-threshold} may represent outliers with respect to  $\mu + 3*stddev$.} As we recall from \fig\ref{fig:Replacement_for_Figure4_CharacterizeReputationFraud}, suspicious reputations can show unusual gain in their growth. Therefore, by looking at these three specific cases, we hoped to find more suspicious users than our baselines $B_{U}$ (i.e., any users showing reputation growth over the average).

\begin{table}[t]
  \centering
  \caption{The different configurations (Case) of Algorithm 2}
    \begin{tabular}{llrl}\toprule
    \textbf{Case} & \textbf{Metric} & \multicolumn{1}{l}{\textbf{Users}} & \textbf{Rationale} \\
    \midrule
    C1    & $\phi_{m,n}^i>28$ & 20,566 & Reputation growth above $\mu+2*stddev$ \\
    C2    & $\phi_{m,n}^i>65$ & 7,380  & Reputation growth  at the 99th percentile \\
    C3    & $\phi_{m,n}^i>130$ & 1,858  & Reputation growth at the 100th percentile \\
    \bottomrule
    \end{tabular}%
  \label{tab:algo2-growth-threshold}%
\end{table}%
\begin{table}[t]
  \centering
  \caption{Performance of Algorithm 2 With Different Configurations}
    \begin{tabular}{lrrrrr}\toprule
    \textbf{Case} & {\textbf{\#User}} & {\textbf{Precision}} & {\textbf{Recall}} & {\textbf{F1-Score}} & {\textbf{Accuracy}} \\
    \midrule
    C1 & 84    & 0.61  & 0.84  & 0.70   & 0.74 \\ 
    C2 & 87    & 0.86  & 0.88  & 0.87  & 0.87 \\ 
    C3 & 91    & 0.87  & 0.89  & 0.88  & 0.87 \\ 
    \midrule
    Total/Avg & 262    & 0.78  & 0.87  & 0.82  & 0.83 \\ 
    \bottomrule
    \end{tabular}%
  \label{tab:algo2-performance}%
\end{table}%

In \tbl\ref{tab:algo2-performance}, we show the performance of the three cases. The performance of algorithm 2 increases as we look at users with more unusual reputation growth, i.e., 
case C3 (i.e., growth more than 130 times the average) shows the best precision, followed by C2 (growth more than 65 times), and then C1 (growth more than 28 times). Both C2 and C3 show 
better performance than the six cases we analyzed for algorithm 1. All the cases show better performance than the baselines. 
For all the cases, the validated reputation frauds on average have much more reputation scores than the non-validated frauds 
(i.e., the ones our algorithm considered as suspicious, but Stack Overflow did not). 
For C1, the average score of the validated frauds is 6,224 and 2,867 for the non-validated frauds (ratio = 2.2). 
This ratio stays the same  (with higher scores) in the other two cases (2.0 for C2 and 2.3 for C3). The higher score of the validated frauds could indicate that for Stack Overflow 
those frauds are consistently more aggressive and/or older than the lower scored non-validated frauds, i.e., Stack Overflow fraud detection system may 
have a system in place to monitor higher scored users that show unusual growth.   
\revision{The numbers of users in each case of algorithm 2 are presented in \tbl\ref{tab:algo2-growth-threshold}. To calculate Precision, Recall, F1-Score, and Accuracy for each case, we have randomly selected several users from each case which is shown in \tbl\ref{tab:algo2-performance}. As previously mentioned, case C3 is more likely to contain suspicious users compared to C2, and C2 is more likely to have more suspicious users than C1.  At first, we selected 84 users for each case (C1, C2, and C3) to evaluate the metrics. 
Although we expected a significantly higher percentage of detection in C3 compared to C2 and C1, our analysis showed that C2 performed much better than C1, while C3 demonstrated a performance almost similar to that of C2. In other words, an increase in reputation jumps beyond a certain threshold did not yield significant differences. To validate our findings, we included additional sample users for C2 and C3 and recalculated the metrics (Precision, Recall, F1-Score, and Accuracy). Surprisingly, the numbers once again indicated that substantial reputation jumps did not significantly enhance the detection of suspicious users. In summary, while reputation jumps do impact the identification of suspicious activities, their influence appears to level off after reaching a certain threshold.}


\begin{table}[t]
  \centering
  \caption{Basic statistics of detected suspicious users' reputation based on various runs of different versions of suspicious user detection based on Algorithm 2}
  \begin{tabular}{llllllll}
    \toprule
    \rotatebox[origin=c]{0}{\bf{Algorithm}} &
    \rotatebox[origin=c]{0}{\bf{Case}} &
    \rotatebox[origin=c]{0}{\bf{Ave(Rep) $\pm$ S.D.}} &
    \rotatebox[origin=c]{0}{\bf{Median(Rep)}} &
    \rotatebox[origin=c]{0}{\bf{Variance(Rep)}} &
    \rotatebox[origin=c]{0}{\bf{Min(Rep)}} &
    \rotatebox[origin=c]{0}{\bf{Max(Rep)}} \\
    \midrule
    Algorithm 2 & C1 & $4695.21 \pm 6839.18$ & 2352 & 46774425.78 & 477 & 39962 \\
    Algorithm 2 & C2 & $23883.16 \pm 25783.95$ & 14820 & 664812133 & 1454 & 145906 \\
    Algorithm 2 & C3 & $18003.79 \pm 20732.18$ & 12785 & 429823143.9 & 2199 & 158687 \\
    \bottomrule
  \end{tabular}
  \label{tab:detailed-statistics-of-userRep-Algo2}
\end{table}

\begin{table}[t]
  \centering
  \caption{Statistics of reduced reputations of users detected by Algorithm 1}
  \begin{tabular}{llllllll}
    \toprule
    \rotatebox[origin=c]{0}{\bf{Algorithm}} &
    \rotatebox[origin=c]{0}{\bf{Case}} &
    \rotatebox[origin=c]{0}{\bf{Ave(Rep) $\pm$ S.D.}} &
    \rotatebox[origin=c]{0}{\bf{Median(Rep)}} &
    \rotatebox[origin=c]{0}{\bf{Variance(Rep)}} &
    \rotatebox[origin=c]{0}{\bf{Min(Rep)}} &
    \rotatebox[origin=c]{0}{\bf{Max(Rep)}} \\
    \midrule
    Algorithm 1: $GC_{V2}$ & C4  & $-203.34 \pm 369.34$ & -27.5 & 136412.77 & -5 & -1950 \\
    Algorithm 1: $GC_{V1}$ & C9  & $-211.86 \pm 440.87$ & -49 & 194365.29 & -5 & -3212 \\
    Algorithm 1: $GC_{V3}$ & C14 & $-98.86 \pm 265.41$ & 0 & 70441.92 & -2 & -1615 \\
    Algorithm 1: $GC_{V3}$ & C16 & $-116.48 \pm 272.25$ & 0 & 74120.4 & -5 & -1615 \\
    Algorithm 1: $GC_{V3}$ & C18 & $-240.88 \pm 370.71$ & -70 & 137423.72 & -5 & -1227 \\
    Algorithm 1: $GC_{V3}$ & C20 & $-200.18 \pm 354.23$ & -5 & 125476.74 & -10 & -1227 \\
    \bottomrule
  \end{tabular}
  \label{tab:detailed-statistics-of-reduced-userRep-Algo1}
\end{table}

\begin{table}[t]
  \centering
  \caption{Statistics of reduced reputations of users detected by Algorithm 2}
  \begin{tabular}{llllllll}
    \toprule
    \rotatebox[origin=c]{0}{\bf{Algorithm}} &
    \rotatebox[origin=c]{0}{\bf{Case}} &
    \rotatebox[origin=c]{0}{\bf{Ave(Rep) $\pm$ S.D.}} &
    \rotatebox[origin=c]{0}{\bf{Median(Rep)}} &
    \rotatebox[origin=c]{0}{\bf{Variance(Rep)}} &
    \rotatebox[origin=c]{0}{\bf{Min(Rep)}} &
    \rotatebox[origin=c]{0}{\bf{Max(Rep)}} \\
    \midrule
    Algorithm 2 & C1 & $-125.36 \pm 278.07$ & -20 & 77323.68 & -2 & -1388 \\
    Algorithm 2 & C2 & $-205.57 \pm 296.93$ & -90 & 88169.83 & -2 & -1670 \\
    Algorithm 2 & C3 & $-211.53 \pm 333.95$ & -80 & 111521.81 & -2 & -1516 \\
    \bottomrule
  \end{tabular}
  \label{tab:detailed-statistics-of-reduced-userRep-Algo2}
\end{table}

\revision{In \tbl\ref{tab:detailed-statistics-of-userRep-Algo2}, we presented more statistics about users' reputations in the three cases of algorithm 2 presented in \tbl\ref{tab:algo2-growth-threshold} of the paper.}

\subsubsection{A discussion on the reduced reputations of users detected by algorithm 1 and algorithm 2 }\label{subsec:reduced-reputation-algo1-and-algo2}
\revision{We do not know whether some people gain reputations just to unlock some features or not. 
We have looked closely into how many scores were removed from the reputation history of users detected by our algorithms.  Two forms of reduction in reputation history occur that are either tagged `reversal' or `removed'.  It is hard to tell whether all the scores are reduced due to fraudulent activities. In the case of `reversal', we are somehow certain that fraud occurred.  On the contrary, `removed' reductions are due to removing the user who caused the reputation gain. Since user removal may be voluntary or due to irresponsible actions and forced by SO.  We are not certain that `removed' reductions are all based on fraud. On the same token, certainly not every removal is fraudulent.  We just looked into scores removed by Stack Overflow without complete knowledge of why they are removed. Based on this assumption, we can estimate the removals as an indication of what score results were targeted by the users in the gaming scenarios. The estimate is rough since more complicated and advanced gaming may go undetected or our estimate may be indeed by overestimate of the target users were seeking. 
\tbl\ref{tab:detailed-statistics-of-reduced-userRep-Algo1} shows some statistics about reduced reputations of users detected by algorithm 1 and \tbl\ref{tab:detailed-statistics-of-reduced-userRep-Algo2} shows some statistics about reduced reputations of users detected by algorithm 2.} 

\begin{tcolorbox}[opacityback=0, standard jigsaw, title=Summary of Study 2 \hrule
\it{Detection of suspicious reputations}]
We propose two algorithms to automatically detect suspicious reputations in Stack Overflow: \begin{inparaenum}[(1)]
\item \bf{Suspicious Community.} We detect suspicious communities that are isolated and whose members have a lot of communications with each other, accept each other's answers repeatedly, communicate with each other often in a very short time, and share similar contents. 
\item \bf{Suspicious User.} We detect suspicious users who show \it{unusual} jump in their reputation scores in a very short time. 
\end{inparaenum} The output of each algorithm is a list of potentially suspicious users. We evaluated the algorithms using 9 different configurations and against two baselines.  
The algorithms show precision 0.41-0.87 and recalls 0.50-0.89 (i.e., F1-score = 0.53-0.88). The algorithms outperformed both baselines (F1-score = 0.12-0.43).
\end{tcolorbox}

\section{Discussion}\label{sec:discussion}

\subsection{\revision{Feedback from Stack Overflow}}\label{subsec:discussion_Feedback_from_StackOverflow}
\begin{figure*}[tp]
\centering
	\centering

   \includegraphics[width=0.9\textwidth,height=\textheight,keepaspectratio]{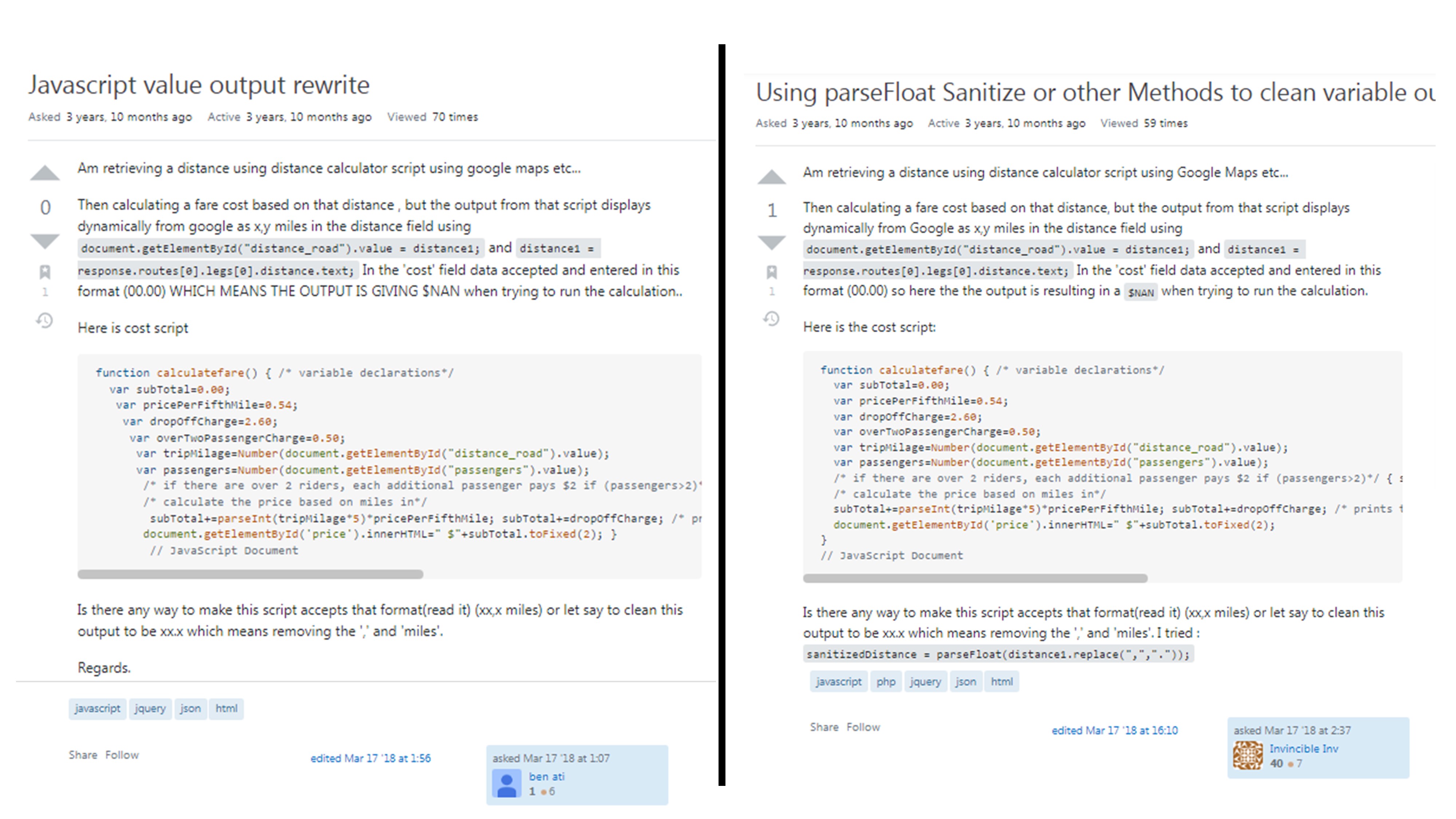}
   	\caption{Example of similar contents of questions which were posted by users detected by algorithm $GC_{V3}$}
   	 \label{fig:Example_of_similar_contents_of_questions}
\end{figure*}

\revision{We reached out to the Stack Overflow team and had a meeting in which we shared our findings and received their feedback. The team acknowledged all the cases that were presented to them and also made suggestions that could help further improve the detection of fraudulent activities, for example by analyzing the distribution of users' activities at large on the platform and using it to detect anomalous activities. They also pointed out their current ongoing efforts (https://charcoal-se.org/) to increase data analytics on the platform, aiming for a better understanding of the activities occurring there. Moreover, they mentioned some of their detection algorithms are not completely automated. \fig\ref{fig:Example_of_similar_contents_of_questions} shows an example of similar contents of two questions that were posted by users detected by our algorithm $GC_{V3}$ and it was acknowledged by the Stack Overflow platform management team.}

\subsection{\revision{Implications of Findings}}\label{subsec:discussion_Implications_of_Findings}
We discuss the implications of our findings for three major stakeholders: \begin{inparaenum}[(1)]
\item Stack Overflow \bf{Forum Designers} who develop and maintain the reputation system, 
\item \bf{Forum Users} who rely on the Q\&A platform for their development needs, and 
\item \bf{Research} in software engineering that leverages Stack Overflow contents.    
\end{inparaenum}

\nd\bf{$\bullet$ Forum Designers.} Our proposed algorithm 2 to detect  suspicious reputations mainly focuses on the gain in reputation scores. This decision is based on 
the observation that suspicious users show rapid and unusual gain in score in a short time. This could mean that while some of their exploitation may be 
penalized by Stack Overflow, they may get away with other exploitation. One of our baselines specifically focused on users who showed a reduced score in the most dataset compared to the previous dataset. We find that most of such users dropped scores due to offering bounties, i.e., they awarded other users their scores as rewards to answering a question. 
In fact, between December and September 2018, only 28,960 active users showed a drop in reputation scores but 461,628 users showed an increase in their scores. Thus, 
it is likely that suspicious users have got their scores increased between two dumps. As we noted in our evaluation, both of our proposed algorithms show good performance to detect suspicious reputations in Stack Overflow. Many of those users from both algorithms 
were also assessed by Stack Overflow as probable frauds. As we noted in our evaluation, quite a few of 
the detected users in our algorithms were not considered as suspicious by Stack Overflow. From a suspicious community perspective, if one or more user in a community is found as 
suspicious, the other users in the community could also be suspicious. Therefore, Stack Overflow could leverage our suspicious community detection algorithm along with its current detection system.

We proposed two algorithms to focus on two types of possible reputation frauds, those who may exploit the reputation 
mechanism in Stack Overflow by forming one or more communities (i.e., mostly by voting ring scenarios), and those who exploit by using diverse other 
scenarios (including voting ring scenarios). This separation was evident in the list of users between the two algorithms in our evaluation. Among all the users 
we analyzed in our evaluation, no user detected by Algorithm 1 was also detected by Algorithm 2, and vice versa. In fact, the  suspicious reputation identified by the two algorithms 
show distinctly different characteristics. The probable validated frauds (i.e., those that were also detected by Stack Overflow) using Algorithm 1 show much less 
reputation scores than those we detected using Algorithm 2. 
When we look at the total scores removed by each such probable validated frauds between two algorithms, the suspicious frauds detected by Algorithm 1  
had on average more scores removed than those probable validated frauds found by Algorithm 2. 
Intuitively, this denotes that suspicious frauds with low reputation scores tend to be more aggressive to increase their scores via exploitation and thus are penalized more than the probable frauds with high scores. 
Given that the suspicious users with high reputation scores tend to not form any specific community, their detection 
could be more challenging. Therefore, both of the algorithms can be used to guide the detection of reputation frauds in Stack Overflow.    

The success of a crowd-source forum depends largely on the active participation of the users. Therefore, 
Stack Overflow may decide to remove/ban a suspicious user from the platform only in the extreme cases. In fact, many of the probable frauds (as detected by both our algorithms 
and Stack Overflow) got their scores reduced for possible exploitation multiple times, yet they are still active in Stack Overflow. Interestingly, while our two algorithms 
found possible frauds with significantly different reputation scores, both these types of suspicious frauds exhibited similar exploitation patterns. For example, on average 
a probable validated fraud user found by Algorithm 1 was engaged in 5.47 suspicious activities, which it is 6.95 for the probable validated users using Algorithm 2. The difference is not 
statistically significant ($p-value = 0.24$). In fact, many of these suspicious frauds may repeatedly exploited the reputation mechanism, e.g., maximum 78 times. 
Given that such suspicious activities are allowed to continue and some users repeatedly exploited the system, the reputation mechanism in Stack Overflow may need to adopt a more 
radical approach to redesign the overall reputation mechanism system.


\nd\bf{$\bullet$ Forum Users.} The reputation frauds can harm the forum content quality by posting low quality and similar posts (recall from our Algorithm 1). 
This can then erode the trust of good users on the overall quality of the crowd-sourced platform. In the long run, the forums become less useful with more lower quality
contents. Such concern is already expressed by a developer in Stack Overflow, \emt{I remember at the beginning of StackOverflow learning a lot by looking up
at the questions. Now... well, finding a good question in the most
popular tags is just hard. I keep trudging through crap.} \bf{MSO$_{A,256304}$}. While concerns about the posted contents is worrying, 
more alarming could be the lack of trust that is brewing in the Stack Overflow user community now due to the prevalence of reputation 
fraud scenarios. We observed large number of users detected by our Algorithm 2 had their reputation scores removed by Stack Overflow. 
Privileges control what you can do on Stack Overflow. users with high reputation scores can gain more privileges. 
Thus, a fraud user with high reputation can promote other frauds to 
influence the overall future direction of the forums.
Indeed, we observed that an admin in Stack Overflow was identified as
using bounty unfairly to promote his friends into the admin level, \emt{The bounties were place on three questions by a high rep user to give a
lower rep user the reputation by rewarding existing answers.
This was to assist the lower rep user attain a rep level that
would give that user moderation privileges.} \bf{MSO$_{A,318504}$}. Indeed, while admins in Stack Overflow are supposed to be trusted for policing the overall 
wellness of the forum, we observed the fraud scenarios can fuel a
lack of trust among developers and forum admins, \emt{\ldots it's so bad you don't trust specific moderators to act in a way
that best represents the community they were elected to serve.} \bf{MSE$_{C,333446}$}.

Two of our observed fraud scenarios involved revenge voting. The frauds can discourage developers' engagement by
promoting bad contents over good contents. A developer, subjected to revenge
voting, was concerned that his posts may not appear in the search results,
due to low/negative scores, \emt{If I "cross" 40 users dumb enough to downvote 2 or 3 of my questions,
my questions will have a 0 score. That is harmful because it would be
like none voted for them, so they won't stand out in searches, and the
content will be less read.} \bf{MSO$_{C,510837}$}
One such impacted user asked Stack Overflow, \emt{I got some mini-revenge (2 or 3 downvotes simultaneously on unrelated
well-written questions) on specific questions (score was 0 or 1, now it's -1 or
0). It has greatly impacted the visibility of those questions. What should I do?}

The prevalence of fraud scenarios and the growing influence of the reputation frauds will increase the lack of trust of good users 
towards the platform, the shared knowledge, and among the users. Unless properly tamed, this lack of trust then can drive away the developers from using Stack Overflow eventually.

\nd\bf{$\bullet$ Researchers.} In recent years, Stack Overflow has become increasingly popular among software developers due to the 
shortcomings in official documentation and alternative knowledge sharing platforms~\cite{Ponzanelli-PrompterRecommender-EMSE2014,Uddin-HowAPIDocumentationFails-IEEESW2015,Uddin-SurveyOpinion-TSE2019}. 
As such, Stack Overflow contents have been extensively leveraged to develop content recommendation system specific to a development need~\cite{Ponzanelli-PrompterRecommender-EMSE2014}, to 
complement API official documentation with usage examples from Stack Overflow~\cite{Subramanian-LiveAPIDocumentation-ICSE2014}, to guide the selection of APIs by 
collecting and summarizing reviews (i.e., positive and negative opinions) about APIs~\cite{Uddin-OpinionValue-TSE2019}, and so on. Given that reputation frauds 
can frequently post low quality contents in Stack Overflow, all of the above research can be impacted. Given that reputation frauds can also downvote to exercise revenge, 
the good quality posts may get lower scores and thus may not be picked in one search engine results or any customized search-based research leveraging Stack Overflow (e.g., AnsweBot~\cite{BowenXu-AnswerBot-ASE2017}). 
Our research findings thus call for an increased awareness towards the prevalence of reputation gaming scenarios in software engineering research that leverage Stack Overflow data to guide 
important development decisions.

\subsection{\acceptance{The Severity of The Problem}}\label{subsec:The_Severity_of_The_Problem}
\acceptance{We provide a rough estimate to quantitatively illustrate the scale of the problem. 
To find out the percentages of fraudulent activities and users among all the Stack Overflow posts, we can employ several approaches.
One estimate can be based on the extrapolation of results of randomly selected 84 users to calculate recall.  When we looked closely at their reputation history, we noticed four had `reversal' and ten had `reversal' or `removed'.  `Reversal' is an indication of reputation decrease due to reputation gaming. In the interview we had with the Stack Overflow team, we noticed the Stack Overflow team looks at `Reversal' as a sign that Stack Overflow has already detected gaming activity.  However, `removed' can have both normal and abnormal activity.   Thus, it can be estimated that 4\% of users have definitively engaged in suspicious activities.
Since we have several fraud scenarios, assuming that the number of users who have answered each other's questions in a round-robin fashion can be used as a rough estimate of all fraudulent activities altogether.   For sure, not all the users who answer other users who previously responded to their questions are conducting improper activities, so the calculated number is an overestimate of voting ring scenarios. On the same note, various other forms of fraudulent and suspicious activities exist, which have not been accounted for. Thus, we can assume that just looking at the overall suspected number of users in voting-ring may be a rough estimate of overall fraud activities within the community to estimate the overall scale of the problem. 
Among 9,321,924 users in SOTorrent dataset September 2018~\cite{website:sotorrent-23092018}, 21,072 of them were involved in round-robin answering and questioning each other's posts.  That is, roughly \(\frac{21,072 }{9,321,924}= 0.23 \%\) of users are involved in suspicious activities.
As a result, we could say between 0.23\% to 4\% of users in stack overflow had some form of gaming activities.
}
\section{Threats to Validity}\label{sec:threats}
\nd\bf{$\bullet$ \revision{External Validity}}
\revision{threats relate to the generalizability of our findings. We analyzed posts returned by the search queries "voting+ring", "moderator+voting+ring", "manipulation", and "reputation fraud". Although other relevant queries (e.g., bounty+gaming) could return additional posts, further exploration could be done. However, we believe our results already cover valid fraud patterns that could be further validated by running new related studies. Furthermore, new studies exploring web forums in StackExchange or not, with similar reputations, could be performed, and possibly reporting new fraud patterns. Also, according to Algorithm 2, we focused on examining consecutive data dumps from Stack Overflow. We chose a three-month interval because that is typically how long it takes for Stack Overflow to release new dumps. Evaluating different time windows could show different results, but they may also complement our current findings. While we do not anticipate different outcomes from analyzing data from other Stack Overflow data dumps, it could still complement our existing results.}


\nd\bf{$\bullet$ Internal Validity} threats relate to experimenter bias and errors while conducting the analysis and while 
reporting the results. The accuracy of analysis relies on the content labeling and the various configuration parameters of our proposed algorithms.
The inherent ambiguity in labeling introduces a
threat to investigator bias. To mitigate the bias,
we conducted reliability analysis.
We observed high agreement level between the coders. To avoid bias in the evaluation of our algorithms, we experimented with different 
configuration (i.e., threshold values) and report the performance of the algorithms under multiple settings. Our ground truth for evaluation is obtained 
from Stack Overflow. We are not aware of how Stack Overflow detects the reputation frauds, neither does Stack Overflow know 
of our approach. This complete lack of awareness mitigates bias in designing the algorithms. 
Thus the agreement between our algorithm and Stack Overflow offers increased confidence on the validity of our finding. In assessing our findings, we only focused on reputation score removal types of `removed' and `reversal' as a mean to assess the correctness of our algorithm.  It should be noted that `removed' reputation score removal may seldom be based on users own requests.  Thus, we used `probable validated fraud' instead of being completely sure on findings.

\nd\bf{$\bullet$ Construct Validity} threats relate to the difficulty in finding data relevant to identify reputation gaming scenarios and 
reputation gamers. We use the personalized reputation dashboard of Stack Overflow users, where Stack Overflow shares each activity of a user that 
contributed to a gain/loss in reputation score of the user. We only consider two `activities', where Stack Overflow decided to remove the score of a user 
due to the manipulation of reputation mechanism of the user. It is possible that Stack Overflow may have missed a reputation exploitation of a user.   



\section{Related Work}\label{sec:related-work}
 Related work can broadly be divided into two areas: \begin{inparaenum}
 \item Understanding the impact of reputation systems, and  
\item Gaming of reputation systems in online forums. 
 \end{inparaenum}

\subsection{Understanding the Impact of Reputation Systems}  
As noted in \sec\ref{sec:introduction}, in traditional crowd-sourced infromation-sharing venues, the role of a knowledge provider/seeker can be different than 
a developer participating in an open source software project. While an open source developer contributes 
for a direct need/support/enjoyment of building a software, 
a knowledge provider/seeker is motivated by the sharing of problems with others and in the process 
wishes to learn from others or to become part of a community~\cite{Lakhani-FreeUserToUserAssistance-JRP2003,Vasilescu-SocialQAKnowledgeSharing-CSCW2014,Bird-CollaborationOSS-ICSM2011a,Bird-LatentSocialOSS-FSE2008a}. 

As StackExchange network of sites are becoming popular among developers, 
various studies have offered insights into the motivation of developers to participate. One aspect of the motivation, 
as noted above, is to be able to learn from each other~\cite{Lee-OpenSourceSignallingDevice-WFFA2003}. 
Other motivating factors include the desire to be recognized among peers and by the 
hiring organizations for career advancement~\cite{Capiluppi-AssessingTechnicalCandidatesSocialWeb-IEEESW2013,website:Bethany-sohiring-blog2013}. 
Such recognition is aided through the \it{reputation/incentive} system~\cite{Vasilescu-SocialQAKnowledgeSharing-CSCW2014,website:SO-ReputationSystem-2019}, which 
is adopted from the concepts of \it{gamification}, i.e., using elements from game design to a 
non-game context (i.e., knowledge sharing)~\cite{Deterding-GamificationDesigningForMotivation-Interactions2012,Deterding-GamificationUsingDesignElemetsInNonGaming-CHI2011}. 
Gamification is widely used in online platforms, and it is found to motivate the users to contribute more~\cite{Lakhani-FreeUserToUserAssistance-JRP2003,Anderson-DiscoveringValueCommunityActivitySO-SIGKDD2012,Antin-BadgeInSocialMedia-CHI2011,Cavusoglu-GamificationMotivateVoluntaryCont-CSCW2015}. 

As we noted in \sec\ref{sec:background}, the reputation of a user in Stack Overflow is a function of three elements: score, badge, and status. 
Among the three aspects, the impact of the badge system was studied by a number of research. The findings show that the badge system increases user participation and the participation level of a user is more when he is close to getting a new badge~\cite{Anderson-SteeringUserBehaviorWithBadges-WWW2013}. The same 
observation was also corroborated by Grant et al.~\cite{Grant-EncourageUserBehaviorWithAchievements-MSR2013}. 
In fact, the badge system is found to motivate 
five psychological functions of users: goal setting, instruction, reputation, status/affirmation, and group identification~\cite{Antin-BadgeInSocialMedia-CHI2011}. 
As previously reported by Hsieh et al.~\cite{Hsieh-WhyPayQA-CHI2010}, these pyschological aspects of feeling important/reputed 
can become more influential for user participations than monentary incentives over time in the social media. 

Several studies specifically looked at bounties in Stack Overflow and found offering bounties to be useful 
to reduce response time, to attract more traffic, as well as to increase the likelihood of getting an answer 
to a post~\cite{Nakasai-AreDonationBadgesAppealing-IEEESoftware2019,Zhou-BountiesStackOverflow-EMSE2019,Kanda-TowardsOpenSourceBounty-SANER2017}. A previous study 
by Krishnamurthy and Tripathi~\cite{Krishnamurthy-BountiesInOpenSource-BookChapter2006} in Free/Libre/Open source software (FLOSS) found that 
the response from bounty hunters depends not only on the offered bounty, but also on the workload, the probability of winning the bounty, and value of the bounty. 
In a separate study, Zhao et al.~\cite{Zhao-DeviseEffectivePolicyBugBounty-JIP2017} also found that the diversity of bounty hunters in vulnerability discovering systems can increase the 
producitivity of the dicovery process. For such vulnerability discovery process, bounties are found to be more effective than hiring professional security researchers~\cite{Finifter-EmpiricalStudyVulnerabilityRewards-USENIX2013}. 
However, the correlation between the common vulnerability score system (CVSS) and the awarded bounty was found to be negative~\cite{Munaiah-VulnerabilitySecurityScoringBounty-SWAN2016}, i.e., 
the other factors were also important. Similarity, heterogeneity of the bug bounty contributors was also important~\cite{Hata-HeterogeneityOfContributorsBounty-ESEM2017}, as well as the 
assessment of the market conditions while offering a bounty~\cite{Zhao-DeviseEffectivePolicyBugBounty-JIP2017}.  
Recently, Wang et al.~\cite{Wang-UnderstandingFactorsForFastAnswers-EMSE2017} 
examined the factors impacting fast answers in Stack Overflow. They suggest to improve the reputation system in Stack Overflow to make non-frequent users more active.

Our study complements the above work by offering, for the first time, a comprehensive overview of the type of reputation fraud scenarios that are prevalent in Stack Overflow and by designing algorithms to detect possible reputation gaming automatically. Our study shows that, while the decentralization of crowd-source forums like Stack Overflow are 
becoming increasingly common to promote knowledge sharing, this approach is vulnerable to exploitation by reputation frauds who devise ways to illegally increase their 
reputation scores. Thus, we need studies from gamification theory to understand the behavior of these reputation frauds and to determine ways to discourage them. Otherwise, 
as we noted in \sec\ref{subsec:discussion_Implications_of_Findings}, such unauthorized gaming of the system can impact any techniques that leverage Stack Overflow data, such as, to recommend quality 
posts~\cite{Ponzanelli-ClassifyQualityForumQuestion-QSIC2014,Ponzanelli-ImproveLowQualityPostDetect-ICSME2014,Ya-DetectHighQualityPosts-JIS2015,
Harper-PredictorAnswerQuality-CHI2008,Li-AnswerQualityPredictions-WWW2015,Calefato-HowToAskForTechnicalHelp-IST2018}, 
to produce software documentation~\cite{Subramanian-LiveAPIDocumentation-ICSE2014} and 
to collect and summarize developers' reviews~\cite{Uddin-OpinionValue-TSE2019,Uddin-OpinerReviewAlgo-ASE2017,Uddin-OpinerReviewAlgo-ASE2017}.

\subsection{Gaming of Reputation Systems}
Jan et al.~\cite{Jan-TowardsMonetaryIncentives-Arxiv2017} examined the impact of financial incentives on different players 
in social Q\&A systems. They found that 
financial incentives attract faster response from the experts, but the incentives also drive users to agressively game the system for profits. 
Zhou et al.~\cite{Zhou-BountiesInOpenSource-Arxiv2019} 
looked at bounties offered as monetary incentives in the BountySource platform. They found that users offering financial bounties on long standing issues may risk financial loss. High reputation scores correlate with increased trust among users (e.g., buyers and sellers), as previously found in eBay by Resnick and 
Zeckhauser~\cite{Resnick-TrustEbayRepSystem-BookChapter2002}. Ye et al.~\cite{Ye-StrategicBehaviorReputationSystemEbay-MIS2014} found that when eBay decided to remove reputation history of all sellers, the sellers with previouly low reputation 
scores posed themselves as high quality sellers and manipulated the buyers. As we noted in \sec\ref{subsec:discussion_Implications_of_Findings}, this observation can serve warning to Stack Overflow for users with undetected 
fraud behavior.
 
Wang et al.~\cite{Wang-HowDoUsersReviseAnswers-TSE2018} investigated how developers in Stack Overflow edit 
posts. They found that the frequency of edits of a user increases as he nears getting a new badge. They also found that the number of edits rejected increases as the users makes more edits in a short time. Such editing behavior may introduce low quality contents.  In this paper, we focus on unauthorized gaming in Stack Overflow. These fraud users do not care about content quality 
and thus flood the site with low quality contents. 
%
%

While our algorithms show good performance, we also have false positives according to 
Stack Overflow validation. The change in performance of the algorithms with different thresholds emphasizes the complexity of finding frauds. 
Indeed, decades of research by Dellarocas~\cite{Dellarocas-DesigningReputationSystemsSocialWeb-BostonUni2010} to design reputation systems in social web conclude that 
a complete cure of this problem is almost impossible. However, more studies and algorithms like our studies are warranted to prevent gaming of reputation systems in Stack Overflow. 
Otherwise, the lack of trustworthiness of knowledge shared in Stack Overflow will only escalate~\cite{Uddin-SurveyOpinion-TSE2019,Zhang-AreCodeExamplesInForumReliable-ICSE2018}. 



\section{Conclusions}\label{sec:summary}
\nd\bf{$\bullet$ Summary.} Developer forums encourage participation using a reputation system.
The system can be targeted for abuse. We conducted an empirical study of 1,697 posts from
Stack Overflow meta sites where developers discussed about the different manipulation of reputation events that happened in Stack Overflow. 
Our analysis revealed four reputation gaming scenarios, such as `voting ring' where fraud users form a ring and upvote each other repeatedly. To automatically detect such reputation frauds, we propose two algorithms. 
The first algorithm detects suspicious communities whose members show a high level of interactions among themselves by posting similar contents, responding to each other’s questions in a very short time, and accepting each other's answers. The second algorithm detects suspicious users who 
show unusual jump in their reputation scores in a short time. The output of each algorithm is a list of potentially suspicious users. We evaluated the algorithms using 9 different configurations and against two baselines.  
The algorithms show precision values ranging 0.41-0.87 and recall values ranging 0.50-0.89 (i.e., F1-score = 0.53-0.88). 
The algorithms outperformed both the baselines (F1-score = 0.12-0.43).  We find that not all suspicious users in our detected suspicious communities are punished by Stack Overflow. Therefore, Stack Overflow 
can utilize our proposed algorithms to improve their detection of fraud users. 

\nd\bf{$\bullet$ Implications.} Our findings can guide \begin{inparaenum}
\item forum designers to properly design the reputation system to prevent manipulation, 
\item forum users to be more cautious while reusing knowledge from the forum, and 
\item researchers in software engineering to investigate tools and techniques to guide forum users and designers to detect reputation frauds and scenrarios properly.
\end{inparaenum}

\nd\bf{$\bullet$ Future Work.} 
We are studying the quality and trust attributes that may arise at the intersection of 
user collaboration and shared contents in developer forums. 
Based on the insights gained we will focus on the development of tools and techniques to inform developers of the potential fraud scenarios in the forums and to 
assist relevant stakeholders (e.g., users, hiring managers) in the identification of experts on a subject matter.

\begin{small}
\bibliographystyle{abbrv}
\bibliography{consolidated}
\end{small}

\end{document}